\documentclass[12pt]{article}
\usepackage{euscript,amsmath, amssymb, amsfonts}

\input{tcilatex}

\begin{document}

\title{{\Large Self-adjoint extensions and spectral analysis in Calogero
problem}}
\author{ D.M. Gitman\thanks{%
Institute of Physics, University of Sao Paulo, Brazil; e-mail:
gitman@dfn.if.usp.br}, I.V. Tyutin\thanks{%
Lebedev Physical Institute, Moscow, Russia; e-mail: tyutin@lpi.ru}, and B.L.
Voronov\thanks{%
Lebedev Physical Institute, Moscow, Russia; e-mail: voronov@lpi.ru}}
\date{}
\maketitle

\begin{abstract}
In this paper, we present a mathematically rigorous quantum-mechanical
treatment of a one-dimensional motion of a particle in the Calogero
potential $\alpha x^{-2}$ . Although the problem is quite old and
well-studied, we believe that our consideration, based on a uniform approach
to constructing a correct quantum-mechanical description for systems with
singular potentials and/or boundaries, proposed in our previous works, adds
some new points to its solution. To demonstrate that a consideration of the
Calogero problem requires mathematical accuracy, we discuss some
\textquotedblleft paradoxes\textquotedblright\ inherent in the
\textquotedblleft naive \textquotedblright\ quantum-mechanical treatment. We
study all possible self-adjoint operators (self-adjoint Hamiltonians)
associated with a formal differential expression for the Calogero
Hamiltonian. In addition, we discuss a spontaneous scale-symmetry breaking
associated with self-adjoint extensions. A complete spectral analysis of all
self-adjoint Hamiltonians is presented.
\end{abstract}

\section{Introduction}

In this paper, we present a mathematically rigorous nonrelativistic
quantum-mechanical (QM) treatment of a one-dimensional motion of a particle
in the potential field%
\begin{equation}
V\left( x\right) =\alpha x^{-2},  \label{6.0}
\end{equation}%
singular at the origin. The case of $\alpha >0$ corresponds to repulsion
from the origin; the case of $\alpha <0$ corresponds to attraction to the
origin.

Our aim is twofold. First, although the problem is quite old and
well-studied, both by physicists and by mathematicians (see the discussion
below), we believe that our consideration adds some new points to its
solution. Second, we present another illustration of a uniform approach to
constructing a correct QM description for systems with singular potentials
and/or boundaries, proposed in our previous works \cite{VorGiT06, VorGiT107,
VorGiT207, VorGiT08}. This description incorporates a proper definition of
physical observables as self-adjoint (s.a. in what follows) operators in an
appropriate Hilbert space, see, e.g., \cite{AkhGl81,Naima69,ReeSi72}, with
special attention to possible ambiguities inherent in the description, and
the spectral analysis of the observables. The first example of such a
description, as applied to a relativistic spin-one-half particle moving in
the Coulomb field of arbitrary charge, was presented in \cite{VorGiTC07}.

Starting from the basic papers by Calogero on the exactly solvable
one-dimensional QM models \cite{Calog169,Calog269,Calog71}, the potential (%
\ref{6.0}) is conventionally called the Calogero potential, and the problem
of a QM description of the system of particles with this pair potential is
known as the Calogero problem.

We restrict ourselves to the case of a motion on the semiaxis $\mathbb{R}%
_{+}=[0,\infty )$. The case of the whole axis $\mathbb{R}=(-\infty ,\infty )$%
, or that of a finite interval $[0,a]$, can be considered by the same
methods. We only mention that setting up the corresponding quantum mechanics
(QM) contains more ambiguity.

The Calogero problem on the semiaxis is of physical significance because it
can be considered as the problem of a radial motion of a particle in higher
dimensions in the potential field $V(r)\sim1/r^{2}$; the variable $x$ is
then a radius $r$, cylindrical or spherical. In particular, this problem is
associated with the three-dimensional motion of a charged particle in the
magnetic field of an infinitely thin and infinitely long solenoid, in which
case $x=r$ is the cylindrical radius, or in the field of a magnetic
monopole, in which case $x=r$ is a spherical radius; see, e.g., \cite%
{Tyuti74} and references therein. It is also associated with the
three-dimensional motion of a polarizable atom in the electric field of an
infinitely thin and infinitely long charged wire \cite{AudSkV01}.

The peculiarity of higher-dimensional classical mechanics in the case of
attraction is that under some initial conditions the particle
\textquotedblleft falls to the center \textquotedblright\ in a finite time
interval \cite{LanLi56}, such that the final state at the end of this
interval is a position $\mathbf{r}=0$ and a momentum $\mathbf{p}=\infty $ of
uncertain direction, and the problem arises how to define the motion of the
particle after this time interval. In some sense, QM \textquotedblleft
inherits\textquotedblright\ these difficulties, although gives them a QM
form.

A \textquotedblleft fall to the center\textquotedblright\ manifests itself
in the case of $\alpha <-1/4$ as the unboundedness of the energy spectrum
from below; for example, see \cite{LanLiQM}. In addition, as was found in
the very beginning of QM, the conventional QM methods of finding energy
eigenvalues and eigenfunctions fail in this case \cite{MotMa33}. By
\textquotedblleft conventional\textquotedblright , we mean the customary
methods adopted in physical textbooks and reduced to directly solving the
corresponding differential equations with the only requirements of
square-integrability for bound-state eigenfunctions (discrete spectrum),
local square-integrability at the singularity/boundary, boundeness at
infinity, and \textquotedblleft normalizability to the $\delta $
function\textquotedblright\ for scattering state eigenfunctions (continuous
spectrum). These difficulties are characteristic for all strongly singular
attractive potentials like $V(r)\sim \alpha /r^{n}$, $n\geq 2$, which
initially even raised the doubt whether such potentials fall into the realm
of QM. A possible way out is to declare that strongly attractive potentials
are inadmissible extrapolations of the known physical forces to arbitrarily
small distances and therefore have no physical meaning without cutting off
the singularity. But then the question arises as to what extent a physical
description, in particular, low-energy physics, depends on a cut-off and how
a possible ambiguity in the description can be parametrized. Examining
strongly singular potentials by themselves, we answer this question to some
extent. It is also worth noting that the attractive Coulomb potential is
strongly singular for relativistic particles.

The first step in overcoming the above difficulties was due to Case, who
noted that a quantum Hamiltonian with a strongly singular attractive
potential, in particular, the Calogero potential with $\alpha <-1/4$, is not
defined by the formal differential expression alone, but \textquotedblleft
needs a further specification by requiring a fixed phase for the wave
functions at the origin\textquotedblright , and the phase is
\textquotedblleft an additional (to the functional form of the potential)
parameter\textquotedblright \textrm{\ }\cite{Case50}. This requirement
followed from the orthogonality condition for eigenfunctions of bound states
with different energy eigenvalues. It is remarkable that the phase is not
determined uniquely, so that there exists a one-parameter family of
candidates for the quantum Hamiltonian. A formula for the negative spectrum
of the Calogero Hamiltonian with $\alpha <-1/4$ and an arbitrary fixed phase
was thus first presented. In fact, as we now realize, this was the first
formulation of additional asymptotic s.a. boundary conditions at the
singularity that specified, nonuniquely, an s.a. Hamiltonian, although
self-adjointness (\textquotedblleft hermicity\textquotedblright ) was
understood as the orthogonality and completeness of eigenfunctions, and the
completeness was only declared. The next step was due to Meetz, who pointed
out that a proper treatment of singular potentials, in particular, the
Calogero problem, requires invoking the theory of s.a. extensions of
symmetric operators\footnote{%
To our knowledge, the idea that the mathematical basis for a proper
treatment of QM problems with singular potentils is the theory of s.a.
extensions of symmetric operators goes back to Berezin and Faddeev, who
applied this theory to solving the quantum-mechanical problem with $3$-dim. $%
\delta $-potential \cite{BerFa61}.}, including such notions as deficient
subspaces and deficiency indices \cite{Meetz64}. S.a. Hamiltonians with
singular attractive potentials, and even with some repulsive Calogero
potentials, were then specified in terms of the respective deficient
subspaces, which introduced an extra parameter; the conjecture by Case was
thus confirmed. Proper spectral decompositions of the resulting Hamiltonians
were also systematically elaborated\footnote{%
However, the corresponding analysis was likely to be perceived by physicists
of that time as excessively complicated, and in fact, the experince was
summarily dismissed.}. It was also emphasized that a conventional limiting
cut-off (regularization) procedure does not yield the known correct results.
Since then, many authors have repeatedly returned to the problem of singular
potentials, especially to the Calogero problem, investigating its different
aspects from different standpoints; see, for example, \cite%
{Behnc68,PerPo70,Allil71,FraLaS71,Narnh74,Tyuti74,Haeri78,AGH-KH88,Jacki93,AudSkV01,Shaba02,B-MGG03,FalMuP04}
(the list of references can be significantly extended), including an
elucidation of the physical meaning of a formal procedure of s.a. extensions
and new parameters involved in terms of regularization and \textquotedblleft
renormalization by square-well counterterms\textquotedblright\ \cite%
{BBCKMK01, BawCo03}.

In this paper and the next one, we summarily review all essential
mathematical aspects of the one-particle Calogero problem by using a uniform
approach based on the theory of s.a. extensions of symmetric differential
operators, namely, on a method of specifying s.a. ordinary differential
operators associated with s.a. differential expressions by (asymptotic) s.a.
boundary conditions and on Krein's method of guiding functionals for a
spectral analysis of ordinary s.a. differential operators.

This paper is organized as follows.

To be convinced that a treatment of the Calogero problem requires
mathematical accuracy, we begin the exposition with applying the customary
physical methods outlined above to this problem and discuss some QM
\textquotedblleft paradoxes\textquotedblright\ inherent in such a
\textquotedblleft naive \textquotedblright\ treatment: sec.~2. In sec.~3, we
study all possible s.a. operators (s.a. Hamiltonians) associated with the
formal differential expression for the Calogero Hamiltonian. A complete
spectral analysis of all such s.a. Hamiltonians is given in sec. 3. Here, we
present their spectra and the corresponding complete sets of (generalized)
eigenfunctions. In sec.~4, we discuss spontaneous scale-symmetry breaking
associated with s.a. extensions.

In the next publication, we are going to discuss a new aspect of the
problem, the so-called oscillator representation for the Calogero
Hamiltonians.

\section{A \textquotedblleft naive\textquotedblright\ treatment of the
problem and related paradoxes}

As mentioned above, the consideration of this section is on the so-called
\textquotedblleft physical level of rigor\textquotedblright , or, in other
words, \textquotedblleft naive\textquotedblright , so we actually repeat
here a negative experience of the first researches.

We start with the formal differential expression, or differential operation (%
$d_{x}=d/dx$), 
\begin{equation}
\check{H}=-d_{x}^{2}+\alpha x^{-2},  \label{6a.2}
\end{equation}%
for the Calogero Hamiltonian, and consider it as an s.a. operator $\hat{H}$
in the Hilbert space $\mathfrak{H}=L^{2}\left( \mathbb{R}_{+}\right) $ of
quantum states for any $\alpha $, conventionally without any reservations
about its domain. We say in advance that the latter is precisely the reason
for paradoxes.

In QM, the time evolution governed by an s.a. Hamiltonian $\hat{H}$ is
unitary and is defined for all moments of time, although, as we have
mentioned in Introduction, an analogue of a \textquotedblleft fall to the
center\textquotedblright\ is well-known from textbooks in the case of $%
\alpha <-1/4$: in this case, the spectrum of $\hat{H}$ is unbounded from
below. This is argued \cite{LanLiQM} by considering the singular Calogero
potential as a limit of bounded regularized potentials,%
\begin{equation}
V_{r_{0}}\left( x\right) =\left\{ 
\begin{array}{c}
\alpha x^{-2}\,,\;x\geq r_{0}\,, \\ 
\alpha r_{0}^{-2}\,,\;x<r_{0}\,,%
\end{array}%
\right.  \label{6a.3}
\end{equation}%
with $r_{0}\rightarrow 0$. Indeed, the limit spectrum is not presented;
moreover, an attentive reader can see that there is no limit spectrum, so
that the problem of the spectrum, as well as that of the limit
eigenfunctions for the Calogero Hamiltonian in the case of $\alpha <-1/4$,
remains completely open.

We therefore look at the problem in more detail. It is natural to expect
that in the case of $\alpha \geq 0$ the spectrum of $\hat{H}$ is
nonnegative; the eigenstates are scattering states, and there exist no bound
states, while in the case of $\alpha <0$ we expect a bound state of negative
energy $E_{0}<0$ in addition to scattering states corresponding to the
nonnegative spectrum.

We now turn to some symmetry arguments. It seems evident that the Calogero
Hamiltonian has the scale symmetry: under the scale transformations $%
x\rightarrow x^{\prime }=lx,\ l>0$, the operators $\hat{H}_{0}=d_{x}^{2}$
and $\hat{V}=\alpha x^{-2}$ transform uniformly and are of the same spatial
dimension, $\mathrm{d}_{H_{0}}=\mathrm{d}_{V}=-2$; therefore, the operator $%
\hat{H}$ also transforms uniformly under scale transformations, and $\mathrm{%
d}_{H}=-2$ . This observation is formalized as follows.

We consider the group of scale transformations $x\rightarrow x^{\prime }=lx$%
, $x\in \mathbb{R}_{+}$,\ $\forall l>0$, and its unitary representation in
the space $L^{2}\left( \mathbb{R}_{+}\right) $ of quantum states by unitary
operators $\hat{U}\left( l\right) $,%
\begin{equation}
\hat{U}\left( l\right) \psi \left( x\right) =l^{-1/2}\psi \left(
l^{-1}x\right)  \label{6a.4}
\end{equation}%
(the spatial dimension of wave functions $\psi $ is $\mathrm{d}_{\psi }=-1/2$
because $\left\vert \psi \left( x\right) \right\vert ^{2}$ is the spatial
probability density). The unitarity of $\hat{U}\left( l\right) $ is easily
verified%
\begin{equation*}
\left\Vert \hat{U}\left( l\right) \psi \right\Vert ^{2}=\int_{0}^{+\infty
}dxl^{-1}\left\vert \psi \left( l^{-1}x\right) \right\vert
^{2}=\int_{0}^{+\infty }dx\left\vert \psi \left( x\right) \right\vert
^{2}=\left\Vert \psi \right\Vert ^{2}\,,
\end{equation*}%
as well as the group law $\hat{U}\left( l_{2}\right) \hat{U}\left(
l_{1}\right) =\hat{U}\left( l_{2}l_{1}\right) .$ It is also easily verified
that%
\begin{equation}
\hat{U}^{-1}\left( l\right) \hat{H}\hat{U}\left( l\right) =l^{-2}\hat{H}%
\Longleftrightarrow \hat{H}\hat{U}\left( l\right) =l^{-2}\hat{U}\left(
l\right) \hat{H}\,,  \label{6a.5}
\end{equation}%
or $\mathrm{d}_{H}=-2$.

For completeness, we present an infinitesimal version of scale symmetry. The
unitary scale transformations $\hat{U}\left( l\right) $ can be represented as%
\begin{equation*}
\hat{U}\left( l\right) =\exp \left( i\ln l\hat{D}\right) ,\;\hat{D}%
=ixd_{x}+i/2=-\left( \hat{x}\hat{p}+\hat{p}\hat{x}\right) /2\hbar ,\;\hat{p}%
=-id_{x},
\end{equation*}%
$\hat{D}$ being the s.a. generator of scale transformations. The scale
symmetry algebra for the Hamiltonian $\hat{H}$ is $\left[ \hat{D}\,,\hat{H}%
\right] =-2i\hat{H}\,.$

Let now $\psi _{E}\left( x\right) $ be an eigenfunction of $\hat{H}$ with an
eigenvalue $E$, i.e., $\hat{H}\psi _{E}\left( x\right) =E\psi _{E}\left(
x\right) ,$ then the scale-symmetry operator relation (\ref{6a.5}) applied
to this function yields%
\begin{equation*}
\hat{H}\left[ \hat{U}\left( l\right) \psi _{E}\left( x\right) \right] =l^{-2}%
\hat{U}\left( l\right) \hat{H}\psi _{E}\left( x\right) =\left(
l^{-2}E\right) \hat{U}\left( l\right) \psi _{E}\left( x\right) \,,
\end{equation*}%
which implies that%
\begin{equation*}
\hat{U}\left( l\right) \psi _{E}\left( x\right) =\psi _{l^{-2}E}\left(
x\right) ,\;\forall l>0,
\end{equation*}%
i.e. $\hat{U}\left( l\right) \psi _{E}\left( x\right) $ is an eigenfunction
of $\hat{H}$ with the eigenvalue $l^{-2}E$. But this implies that the group
of scale transformations acts transitively on both the positive and negative
parts of the energy spectrum, so that these parts must either be empty or
occupy the respective positive and negative semiaxis of the real axis.

This is completely consistent with what we expect for the spectrum of $\hat {%
H}$ in the case of repulsion, $\alpha>0$, where $E\geq0.$

But in the case of attraction, $\alpha <0$, we meet paradoxes. Indeed, in
this case we expect at least one bound state with a negative level, $E_{0}<0$%
. But if there exists at least one such state, then, according to scale
symmetry, there must be a continuous set of normalizable bound states with
energies $l^{-2}E_{0}$, $\forall l>0$, and the negative part of the spectrum
is the entire negative semiaxis, i.e., a \textquotedblleft fall to the
center\textquotedblright\ occurs for all $\alpha <0.$

This picture is quite unusual and contradictory, because there can be no
continuous set of normalizable eigenstates for any s.a. operator in $%
L^{2}\left( \mathbb{R}_{+}\right) $: it would contradict the fact that $%
L^{2}\left( \mathbb{R}_{+}\right) $ is a separable Hilbert space. Another
surprising fact is that the spectrum of the Calogero Hamiltonian is not
bounded from below for any $\alpha <0$, not only for $\alpha <-1/4$.

The situation becomes even more entangled if we try to find boundstates of $%
\hat{H}$ corresponding to negative energy levels, $E<0$. The corresponding
differential equation for these eigenstates $\psi _{E}\left( x\right) \equiv
\psi _{k}\left( x\right) $ is%
\begin{equation}
\check{H}\psi _{k}\left( x\right) =-k^{2}\psi _{k}\left( x\right)
\,,\;k^{2}=-E>0\,.  \label{6a.8}
\end{equation}%
There are two \textquotedblleft dangerous\textquotedblright\ points for the
square-integrability of $\psi _{k}\left( x\right) $: the infinity, $x=\infty 
$, and the origin, $x=0$, which is a point of singularity of the potential
and a boundary simultaneously.

The behavior of a solution $\psi _{k}\left( x\right) ,$ if it does exist, at
infinity where the potential vanishes is evident: $\psi _{k}\left( x\right)
\simeq c\exp \left( -kx\right) ,\;x\rightarrow \infty \,.$ This behavior,
which manifests the square-integrability of $\psi _{k}\left( x\right) $ at
infinity, must be compatible with the local square-integrability of $\psi
_{k}\left( x\right) $ at the origin. The existence of $\psi _{k}\left(
x\right) $ for a given $k$ is thus defined by its asymptotic behavior at the
origin, which, because of the singularity, coincides with the asymptotic
behavior of the general solution of the homogeneous equation $\check{H}%
y\left( x\right) =0$ at the origin. The general solution of this equation is%
\begin{equation}
y\left( x\right) =\left\{ 
\begin{array}{l}
x^{1/2}\left( c_{1}x^{\varkappa }+c_{2}x^{-\varkappa }\right) ,\,\alpha \neq
-1/4, \\ 
x^{1/2}(c_{1}+c_{2}\ln x),\ \alpha =-1/4,%
\end{array}%
\right. \,,  \label{6a.10}
\end{equation}%
where%
\begin{equation}
\varkappa =\sqrt{1/4+\alpha }=\left\{ 
\begin{array}{l}
\sqrt[+]{1/4+\alpha },\ \alpha \geq -1/4, \\ 
i\sigma ,\ \sigma =\sqrt[+]{|1/4+\alpha |}>0,\ \alpha <-1/4.%
\end{array}%
\right.  \label{7.2.0}
\end{equation}%
We can see that if $-1/4\leq \alpha <0,$ we have $\varkappa <1/2$, and $%
y\left( x\right) \rightarrow 0$ as $x\rightarrow 0,$ so that $\psi
_{k}\left( x\right) $ is certainly square-integrable at the origin
irrespective of $k.$ The same holds true if $\alpha <-1/4$, in which case $%
\varkappa =i\sigma $ and $y\left( x\right) \rightarrow 0$ infinitely
oscillating as $x\rightarrow 0$. This implies that $\psi _{k}\left( x\right) 
$ exists for any $k>0,$ which confirms the previous arguments that the
negative \textquotedblright discrete\textquotedblright\ spectrum is in fact
continuous and occupies all the negative real semiaxis.

Furthermore, both functions $x^{1/2\pm \varkappa }$ are also
square-integrable if $1/2\leq \varkappa <1,$ i.e., if $0\leq \alpha <3/4$,$\ 
$so that there is a continuous set of negative energy levels unbounded from
below for $\alpha =0$ (the case of a free particle) and even for repulsive
potentials, $V\left( x\right) >0.$ A \textquotedblleft fall to the
center\textquotedblright\ for repulsive potentials is quite paradoxical.

We can present the explicit form of $\psi _{k}\left( x\right) .$ By the
substitution%
\begin{equation}
\psi _{k}\left( x\right) =x^{1/2}u_{k}\left( kx\right) ~,  \label{6a.11}
\end{equation}%
we reduce equation (\ref{6a.8}) to the following equation for the function $%
u(z)=u_{k}(kx),\;z=kx$:%
\begin{equation}
u^{\prime \prime }+z^{-1}u-(1+\varkappa ^{2}z^{-2})u=0\,,  \label{6a.12}
\end{equation}%
whose solutions are the Bessel functions of imaginary argument. It follows
that for $\alpha <3/4$ and for any $k>0$ the square-integrable solution of
the eigenvalue problem (\ref{6a.8}) for bound states is given by $\psi
_{k}\left( x\right) =x^{1/2}K_{\varkappa }\left( kx\right) $, where $%
K_{\varkappa }\left( x\right) $ is the so-called McDonald function.

The final remark is that $\psi _{k}\left( x\right) $ remains
square-integrable for complex $k=k_{1}+ik_{2},\;k_{1}>0$, so that the
seemingly s.a. $\hat{H}$ has complex eigenvalues.

These inconsistencies, or paradoxes, manifest that something is wrong with
QM in the case of singular potentials, as well as in the case of boundaries,
or, at least, something is wrong with our previous considerations following
the conventional methods. It appears that we have been too \textquotedblleft
naive\textquotedblright\ in our considerations; strictly speaking, we have
been incorrect, and our arguments have been wrong. The main reason is that
almost all operators involved are unbounded, while for unbounded operators,
in contrast to bounded operators defined everywhere, the algebraic rules,
the notions of self-adjointness, commutativity, and symmetry are nontrivial.

In particular, we actually implicitly adopted that the operator $\hat{H}$
acts (is defined) on the so-called natural domain, which is the set of
square-integrable functions $\psi$ satisfying the only conditions that the
differential operation\emph{\ }$\check{H}$\emph{\ }is applicable to $\psi $
and $\check{H}\psi $ is also square-integrable.

As we can see below, this operator with $\alpha <3/4$ is not s.a..

\section{Self-adjoint Calogero Hamiltonians}

We now proceed with a more rigorous treatment of the Calogero problem on the
semiaxis $\mathbb{R}_{+}$. The first problem to be solved is constructing
and suitably specifying all Hamiltonians associated with the differential
expression (\ref{6a.2}) as s.a. operators in the Hilbert space $\mathfrak{H}%
=L^{2}\left( \mathbb{R}_{+}\right) $ of QM states; the second problem is a
complete spectral analysis of each of the obtained Hamiltonians, and,
finally, resolving the paradoxes discussed in the previous section, in
particular, the paradox concerning the apparent scale symmetry.

In solving the first problem, we follow \cite{VorGiT207,VorGiT08}; we say in
advance that a solution crucially depends on a value of $\alpha$.

We start with an initial symmetric operator $\hat{H}$ associated with an
even s.a. differential expression $\check{H}$ (\ref{6a.2}) and the operator $%
\hat{H}^{+}$ being the adjoint\footnote{%
From now on, we let the same $\hat{H}$ denote a new operator which differs
from the \textquotedblleft naive Hamiltonian\textquotedblright\ $\hat{H}\,$%
\textrm{\ }in the previous subsection and hope that this will not lead to
confusion. In \cite{VorGiT207}, the operators $\hat{H}$ and $H$ $^{+}$ were
respectively denoted by $\hat{H}^{(0)}$ and $H$ $^{\ast }$. We remind the
reader that the self-adjointness of a differential expression is understood
in the sence of Lagrange \cite{VorGiT207}.} of $\hat{H}$. S.a. Hamiltonians $%
\hat{H}_{U}$ are s.a. extensions of the symmetric $\hat{H}$ and
simultaneously s.a. restrictions of the adjoint $\hat{H}^{+}$; the meaning
of the subscript $U$ labelling s.a. extensions becomes clear below.

All the above operators form a chain of inclusions $\hat{H}$ $\subset \hat{H}%
_{U}\subseteq \hat{H}^{+}$ and differ only by their domains in $L^{2}\left( 
\mathbb{R}_{+}\right) $, while their action on the corresponding domains is
given by the same differential expression\footnote{%
Here and elsewhere, we cite our review, where one can find relevant
references.} (\ref{6a.2}); see \cite{VorGiT207}.

When defining these operators in what follows, we therefore cite only their
domains.

The domain $D_{H}$ of the initial symmetric operator $\hat{H}$ is the space $%
\mathcal{D}\left( \mathbb{R}_{+}\right) $ of smooth functions with a compact
support%
\begin{equation}
D_{H}=\mathcal{D}\left( \mathbb{R}_{+}\right) =\left\{ \varphi (x):\varphi
\in \mathbb{C}^{\infty }\left( \mathbb{R}_{+}\right) ,\ \mathrm{supp}%
\,\varphi \subseteq \left[ \alpha ,\beta \right] \subset \mathbb{(}0,\infty 
\mathbb{)}\right\} ,  \label{6b.1}
\end{equation}%
which is dense in $L^{2}\left( \mathbb{R}_{+}\right) $. Such a choice of $%
D_{H}$ is based on a natural supposition that principal restrictions on
functions from $D_{H}$ must be connected only with peculiarities of the
problem, neighborhoods of boundaries and potential singularities. As is
known \cite{VorGiT207}, the domain\footnote{%
In \cite{VorGiT207}, the natural domain was denoted simply by $D_{\ast }$.
The operator $\hat{H}^{+}$ actually coincides with the \textquotedblleft
naive Hamiltonian\textquotedblright\ $\hat{H}\ $in the previous subsection.} 
$D_{H^{+}}$ of the operator adjoint to $\hat{H}$, $\hat{H}^{+}$, is the
so-called natural domain $D_{\ast \check{H}}$ for $\check{H},$%
\begin{equation}
D_{H^{+}}=D_{\ast \check{H}}=\left\{ \psi _{\ast }(x):\psi _{\ast }\,,\psi
_{\ast }^{\prime }\;\text{\textrm{are a.c. in }}\mathbb{R}_{+};\ \psi _{\ast
},\,\check{H}\psi _{\ast }\in L^{2}(\mathbb{R}_{+})\right\} ,  \label{6b.1a}
\end{equation}%
where an abbreviation \textquotedblleft a.c.\textquotedblright\ stands for
\textquotedblleft absolutely continuous\textquotedblright .

S.a. Hamiltonians $\hat{H}_{U}$ are constructed as s.a. restrictions of $%
\hat{H}^{+}\;$based on the quadratic asymmetry form $\Delta _{H^{+}}$ which
is a measure of the asymmetricity of $\hat{H}^{+}$ and is defined by%
\footnote{%
In \cite{VorGiT107}, this form was denoted by $\Delta _{\ast }$.} 
\begin{equation*}
\Delta _{H^{+}}(\psi _{\ast })=(\psi _{\ast },\hat{H}^{+}\psi _{\ast })-(%
\hat{H}^{+}\psi _{\ast },\psi _{\ast }),\;\forall \psi _{\ast }\in D_{H^{+}},
\end{equation*}%
see \cite{VorGiT107}. If $\Delta _{H^{+}}=0$, the operator $\hat{H}^{+}$ is
symmetric and therefore s.a.; the operator $\hat{H}$ is then essentially
s.a.; its deficiency indices are $(0,0)$, and its unique s.a. extension is
precisely $\hat{H}^{+}$; if $\Delta _{H^{+}}\neq 0$, the deficiency indices\
of $\hat{H}$ are nonzero, and the domain $D_{H_{U}\text{ }}$ of an s.a.
operator $\hat{H}_{U}$ is defined as a maximum subspace of $D_{H^{+}}$ where 
$\Delta _{H^{+}}$ vanishes \cite{VorGiT107}. If we follow the general theory
of s.a. extensions of symmetric operators with equal nonzero deficiency
indices $(m,m)$, these subspaces are determined in terms of the deficient
subspaces of the initial symmetric operator $\hat{H}$ and a unitary operator 
$\hat{U}$ relating them\footnote{%
To be more precise, $\hat{U}$ is an isometry; the term \textquotedblleft
unitary\textquotedblright\ is more conventional for the physical literature.}%
, and there is an $m^{2}$-parameter $U(m)$-family $\{\hat{H}_{U}\}\,$of s.a.
extensions , where $U(m)$ is a unitary group \cite{VorGiT107}.

We now remind the reader of the basic points of a method proposed in \cite%
{VorGiT08} for constructing s.a. differential operators $\hat{f}_{U}$
associated with a general ordinary s.a. differential expression $\check{f}$
defined on an interval of the real axis in case the associated initial
symmetric differential operator $\hat{f}$ with the adjoint $\hat{f}^{+}$ has
equal deficiency indices.

In our opinion, an advantage of this method is that it avoids evaluating
deficiency indices and deficient subspaces of the initial symmetric operator
and allows specifying the s.a. operators by explicit s.a. boundary
conditions, which is convenient for a subsequent spectral analysis.

For differential operators, the quadratic asymmetry form $\Delta _{f^{+}}$
is represented in terms of quadratic boundary forms \cite{VorGiT207}. In our
case, where both ends of the semiaxis are singular, this representation is 
\begin{equation*}
\Delta _{H^{+}}\left( \psi _{\ast }\right) =[\psi _{\ast },\psi _{\ast
}](\infty )-[\psi _{\ast },\psi _{\ast }](0),\ [\psi _{\ast },\psi _{\ast
}](0/\infty )=\lim_{x\rightarrow 0/\infty }[\psi _{\ast },\psi _{\ast }](x),
\end{equation*}%
where 
\begin{equation*}
\,[\psi _{\ast },\psi _{\ast }](x)=\overline{\psi _{\ast }^{\prime }(x)}\psi
_{\ast }(x)-\overline{\psi _{\ast }(x)}\psi _{\ast }^{\prime }(x).
\end{equation*}%
The quadratic local forms $[\psi _{\ast },\psi _{\ast }](0/\infty )$ are the
respective left (at the origin) and right (at infinity) boundary forms;
these forms do exist (being finite) and are independent.

Each boundary form, if it is nonzero, is a quadratic form in asymptotic
boundary coefficients (a.b. coefficients) that are the boundary values of
functions $\psi _{\ast }\in D_{f^{+}}$ and their derivatives\footnote{%
In the case of even differential expressions, the derivatives are replaced
by so-called quasiderivatives \cite{VorGiT207}.} if the respective end
(boundary) of the interval is regular or the numerical coefficients in front
of the linearly-independent leading terms defining the asymptotic behavior
of these functions at the respective end and giving a nonzero contribution
to the boundary form if the end is singular. Therefore, the asymmetry form $%
\Delta _{f^{+}}$ is a quadratic in all a.b. coefficients $\{c_{k}\}_{1}^{2m}$%
. Linearly combining the a.b. coefficients into so-called diagonal a.b.
coefficients$\,\{c_{+,k}\}_{1}^{m}$ \ and $\{c_{-,k}\}_{1}^{m}$ of the same
dimension, we reduce this quadratic form to a diagonal canonical form, 
\begin{equation*}
\Delta _{H^{+}}\left( \psi _{\ast }\right) =2i\kappa \left(
\sum_{k=1}^{m}|c_{+,k}|^{2}-\sum_{k=1}^{m}|c_{-,k}|^{2}\right) ,
\end{equation*}%
where $\kappa $ is a real factor. We note that the inertia indices of the
quadratic form coincide with the deficiency indices that are found in
passing when finding the a.b. coefficients.

Any s.a. operator $\hat{f}_{U}$ associated with a given s.a. differential
expression $\check{f}$ is uniquely specified by additional boundary
conditions at the ends of the interval on the functions $\psi _{\ast }\in
D_{f^{+}}$. These boundary conditions are called s.a. boundary conditions;
in the presence of singular ends s.a. boundary conditions are of asymptotic
form and are called asymptotic s.a. boundary conditions (a.b. conditions ).

(Asymptotic) s.a. boundary conditions are defined by a (fixed) unitary $%
m\times m$ matrix $U=\parallel U_{kl}\parallel ,\ k,l=1,...,m$, that
establishes the isometric relation 
\begin{equation}
c_{-,k}=\sum_{k=1}^{m}U_{kl}c_{+,l}  \label{6b.5d}
\end{equation}%
between the diagonal a.b. coefficients and thus define a maximum subspace $%
D_{U}\subseteq D_{f^{+}}$, where the quadratic asymmetry form $\Delta
_{f^{+}} $ vanishes identically.

The subspace $D_{U}$ is the domain of an s.a. operator $\hat{f}_{U}$, $%
D_{U}=D_{f_{U}}$. In case both ends of the interval are regular, relation (%
\ref{6b.5d}) is a relation between the boundary values of functions $\psi
_{U}\in $\ $D_{f_{U}}$ and their derivatives, and defines customary boundary
conditions. In the case of singular ends, relation (\ref{6b.5d}) prescribes
the asymptotic behavior of functions $\psi _{U}\in $\ $D_{f_{U}}$ at the
respective ends; more precisely, the a.b. conditions are formulated as
explicit formulas for the leading asymptotic terms of the functions $\psi
_{U}$ at the respective ends.

Conversely, any unitary $m\times m$ matrix $U$ uniquely defines an
associated s.a. operator $\hat{f}_{U}$ by relation (\ref{6b.5d}), so that
there exists an $m^{2}$-parameter $U(m)$-family $\{\hat{f}_{U}\}$ of s.a.
operators associated with a given s.a. differential expression $\check{f}$.

With this \textquotedblleft instructions\textquotedblright , we return to
constructing s.a. Hamiltonians associated with the Calogero differential
expression $\check{H}$ (\ref{6a.2}); it is natural to use the subscript%
\textrm{\ }$U$, or an equivalent one, for the notation of these operators.

The first step consists in evaluating the boundary forms $[\psi ,_{\ast
},\psi _{\ast }](\infty )$ and $[\psi ,_{\ast },\psi _{\ast }](0)$ in terms
of a.b. coefficients.

Because the Calogero potential $V(x)$ (\ref{6.0}) vanishes at infinity, we
have $\psi _{\ast }(x)$, $\psi _{\ast }^{\prime }(x)\rightarrow 0$ as $%
x\rightarrow \infty $, so that $[\psi _{\ast },\psi _{\ast }](\infty )=0$.
In other words, the infinity turns out to be irrelevant and the asymmetry
form $\Delta _{H^{+}}$ is reduced to the boundary form at the origin and is
given by 
\begin{equation}
\Delta _{H^{+}}\left( \psi _{\ast }\right) =-[\psi _{\ast },\psi _{\ast
}](0).  \label{6b.5e}
\end{equation}

Therefore, constructing an s.a. Hamiltonian associated with an s.a. Calogero
differential expression $\check{H}$ (\ref{6a.2}) is reduced to finding a
maximum subspace in $D_{\ast \check{H}}$ where the boundary form at the
origin vanishes identically. The physical meaning of the latter condition is
clear: because the quadratic local form $\,[\psi _{\ast },\psi _{\ast }](x)$
is the probability flux, up to a numerical factor, its vanishing at the
origin implies that the probability flux at the origin is zero and a
particle does not escape the semiaxis through the left end, together with
the zero probability flux at\textrm{\ }infinity; this implies the unitarity
of the time evolution generated by the Hamiltonian that must be s.a.

An evaluation of the boundary form at the origin requires finding a behavior
of the wave functions $\ \psi _{\ast }\in D_{\ast \check{H}}$ and their
derivatives $\psi _{\ast }^{\prime }$ at the origin. This behavior is
conventionally established as follows. We consider the relation $\check{H}%
\psi _{\ast }=\chi $ as a differential equation with respect to the function 
$\psi _{\ast }$ with a given $\chi \in L^{2}\left( \mathbb{R}_{+}\right) $.
If we omit the condition $\psi _{\ast }\in D_{\ast \check{H}}$ for a while,
the general solution of this equation and its first derivative allow
standard integral representations in terms of the nonhomogeneous term $\chi $
and linearly independent solutions $y_{1}=x^{1/2+\varkappa }$, and $y_{2},$%
\begin{equation*}
\,y_{2}=\left\{ 
\begin{array}{c}
x^{1/2-\varkappa },\;\alpha \neq -1/4\ (\varkappa \neq 0), \\ 
x^{1/2}\ln x,\;\alpha =-1/4\,(\varkappa =0),%
\end{array}%
\right. ,\,
\end{equation*}%
of a homogeneous equation, see (\ref{6a.10}), with the Wronskian%
\begin{equation*}
\mathrm{Wr}\left( y_{1},y_{2}\right) =\left\{ 
\begin{array}{c}
-2\varkappa ,\;\alpha \neq -1/4, \\ 
1,\;\alpha =-1/4,%
\end{array}%
\right.
\end{equation*}%
where $\varkappa $ is given by (\ref{7.2.0}). These representations are 
\begin{align}
\psi _{\ast }(x)& =-\frac{x^{1/2}}{2\varkappa }\left[ x^{\varkappa
}\int_{a}^{x}d\xi \xi ^{1/2-\varkappa }\chi -x^{-\varkappa }\int_{0}^{x}d\xi
\xi ^{1/2+\varkappa }\chi \right]  \notag \\
& +c_{1}\left( k_{0}x\right) ^{1/2+\varkappa }+c_{2}\left( k_{0}x\right)
^{1/2-\varkappa }\,,  \notag \\
\psi _{\ast }^{\prime }(x)& =-\frac{x^{-1/2}}{2\varkappa }\left[ \left(
1/2+\varkappa \right) x^{\varkappa }\int_{a}^{x}d\xi \xi ^{1/2-\varkappa
}\chi -\left( 1/2-\varkappa \right) x^{-\varkappa }\int_{0}^{x}d\xi \xi
^{1/2+\varkappa }\chi \right]  \notag \\
& +[c_{1}\left( k_{0}x\right) ^{1/2+\varkappa }+c_{2}\left( k_{0}x\right)
^{1/2-\varkappa }\,]^{\prime },\;\alpha \neq -1/4\;(\varkappa \neq 0),
\label{6b.7} \\
\psi _{\ast }(x)& =x^{1/2}\left[ \int_{0}^{x}d\xi \xi ^{1/2}\ln (k_{0}\xi
)\chi -\ln (k_{0}x)\int_{0}^{x}d\xi \xi ^{1/2}\chi \right] +  \notag \\
& +c_{1}x^{1/2}+c_{2}x^{1/2}\ln (k_{0}x)\,,  \notag \\
\psi _{\ast }^{\prime }(x)& =x^{-1/2}\left[ \frac{1}{2}\int_{0}^{x}d\xi \xi
^{1/2}\ln (k_{0}\xi )\chi -(1+\frac{1}{2}\ln (k_{0}x))\int_{0}^{x}d\xi \xi
^{1/2}\chi \right]  \notag \\
& +\left[ c_{1}x^{1/2}+c_{2}x^{1/2}\ln (k_{0}x)\,\right] ^{\prime },\;\alpha
=-1/4,\;(\varkappa =0),  \notag
\end{align}%
where $k_{0}$ is an arbitrary, but fixed, parameter of the dimension of
inverse length introduced by dimensional reasons, $a>0$ for $\alpha \geq 3/4$%
, $a=0$ for $\alpha <3/4$, and $c_{1}$ and $c_{2}$ are arbitrary numerical
coefficients. We note that $c_{1,2\text{ }}$ are of the same dimension of
one half of the inverse length, the dimension of functions $\psi _{\ast }$,
whereas the dimension of functions $\chi =\check{H}\psi _{\ast }$ is $5/2$
of the inverse length.

The asymptotic behavior of the integral terms in (\ref{6b.7}) at the origin
is estimated using the Cauchy--Bunyakovskii inequality. For example, if $%
\alpha >-1/4$ $(\varkappa >0)$, the Cauchy--Bunyakovskii inequality yields 
\begin{equation*}
\left\vert x^{1/2-\varkappa }\int_{0}^{x}d\xi \xi ^{1/2+\varkappa }\chi
\right\vert \leq x^{1/2-\varkappa }\left[ \int_{0}^{x}d\xi \xi
^{1+2\varkappa }\right] ^{1/2}\left[ \int_{0}^{x}d\xi \left\vert \chi
\right\vert ^{2}\right] ^{1/2},
\end{equation*}%
and with the estimates%
\begin{equation*}
\left[ \int_{0}^{x}d\xi \xi ^{1+2\varkappa }\right] ^{1/2}=O\left(
x^{1+\varkappa }\right) ,\ \left[ \int_{0}^{x}d\xi \left\vert \chi \left(
\xi \right) \right\vert ^{2}\right] ^{1/2}\rightarrow 0\ \text{as }%
x\rightarrow 0
\end{equation*}%
(the second estimate follows from the fact that $\chi \left( \xi \right) \in
L^{2}(\mathbb{R}_{+})$), we find%
\begin{equation}
\left\vert x^{1/2-\varkappa }\int_{0}^{x}d\xi \xi ^{1/2+\varkappa }\chi
\right\vert =O(x^{3/2})\,.  \label{6.2.ab}
\end{equation}%
The r.h.s. in (\ref{6.2.ab}) is overestimated: $O(x^{3/2})$\ can be replaced
by $o(x^{3/2})$. Similarly, for the integral term $x^{1/2+\varkappa
}\int_{a}^{x}d\xi \xi ^{1/2-\varkappa }\chi (\xi ),\,a>0,$ we obtain that
the the estimates $O(x^{3/2})$ and $O(x^{3/2}\sqrt{\left\vert \ln
x\right\vert })$ hold in the respective cases of $\alpha >3/4$ $(\varkappa
>1)$, and $\alpha =3/4$ $(\varkappa =1)$, while the estimate $O(x^{3/2})$
holds for the sum of both integral terms in the case of $\alpha =-1/4$.

The asymptotic behavior of the free terms in (\ref{6b.7}) is evident. If we
now restore the condition\ $\psi _{\ast }(x)\in L^{2}(\mathbb{R}_{+})$, we
find that $c_{2}\neq 0$ in the case of $\alpha \geq 3/4\,(\varkappa \geq 1)$
contradicts the condition, because in this case the function $c_{2}\left(
k_{0}x\right) ^{1/2-\varkappa }$ is not square-integrable at the origin
unless $c_{2}=0$.

The asymptotic behavior of the derivative $\psi_{\ast}^{\prime}$ at the
origin is established quite similarly.

The estimates for the asymptotic behavior of $\psi _{\ast }$ and $\psi
_{\ast }^{\prime }$ at the origin, i.e., as $\ x\rightarrow 0$, are finally
given by%
\begin{align}
\psi _{\ast }(x)& =\psi _{\ast }^{\mathrm{as}}(x)+\left\{ 
\begin{array}{l}
O(x^{3/2}),\ \alpha \neq 3/4\;(\varkappa \neq 1), \\ 
O(x^{3/2}\sqrt{\left\vert \ln x\right\vert }),\ \alpha =3/4\;(\varkappa =1),%
\end{array}%
\right\vert  \notag \\
\psi _{\ast }^{\prime }(x)& =\psi _{\ast }^{\mathrm{as}}\prime (x)+\left\{ 
\begin{array}{l}
O(x^{1/2}),\ \alpha \neq 3/4\;(\varkappa \neq 1), \\ 
O(x^{1/2}\sqrt{\left\vert \ln x\right\vert }),\ \alpha =3/4\;(\varkappa =1),%
\end{array}%
\right\vert  \label{7b.4a}
\end{align}%
where%
\begin{equation}
\psi _{\ast }^{\mathrm{as}}(x)=\left\{ 
\begin{array}{l}
0,\;\alpha \geq 3/4\ (\varkappa \geq 1), \\ 
c_{1}\left( k_{0}x\right) ^{1/2+\varkappa }+c_{2}\left( k_{0}x\right)
^{1/2-\varkappa },\mathrm{\;}\left\{ 
\begin{array}{c}
-1/4<\alpha <3/4\ (\varkappa \geq 1) \\ 
\alpha <-1/4\;(\varkappa =i\sigma ,\ \sigma >0)%
\end{array}%
\right. , \\ 
c_{1}x^{1/2}+c_{2}x^{1/2}\ln (k_{0}x),\;\alpha =-1/4\ (\varkappa =0).%
\end{array}%
\right.  \label{7b.4b}
\end{equation}

The asymptotic estimates (\ref{7b.4a}) and (\ref{7b.4b}) allow a simple
calculation of the asymmetry form $\Delta _{H^{+}}$, given by (\ref{6b.5e})
in terms of a.b. coefficients, and then an explicit formulation of a.b.
conditions specifying all s.a. Hamiltonians $\hat{H}_{U}$ associated with
the Calogero differential expression $\check{H}$ (\ref{6a.2}) via relation (%
\ref{6b.5d}). It is easy to see that the terms like $O(x^{3/2})$ in $\psi
_{\ast \text{ }}$and $O(x^{1/2})$ in $\psi _{\ast \text{ }}^{\prime }$give
zero contributions to $\Delta _{H^{+}}$, while the coefficients $c_{1}$ and $%
c_{2}$ are precisely a.b. coefficients. The result crucially depends on the
value of $\alpha $. According to (\ref{7b.4b}), four regions of the values
of $\alpha $ are naturally distinguished.

\subsection{First region: $\protect\alpha \geq 3/4\,(\varkappa \geq 1)$}

For this region of $\alpha $, the asymmetry form is evidently trivial, $%
\Delta _{H^{+}}=0.$ This implies that the operator $\hat{H}^{+}$ is
symmetric and therefore s.a., which means that the initial symmetric
operator $\hat{H}$ has only one s.a. extension\footnote{%
In passing, we find that $\hat{H}$ has zero deficiency indices and is
therefore\textrm{\ }essentially s.a..} that is just $\hat{H}^{+}$. In other
words, there exists only one s.a. Hamiltonian $\hat{H}_{1}=\hat{H}^{+}$
associated with the s.a. Calogero differential expression $\check{H}$ (\ref%
{6a.2}) with $\alpha \geq 3/4,$ and it is defined on the natural domain $%
D_{\ast \check{H}}$ (\ref{6b.1a}),

\begin{equation}
D_{H_{1}}=D_{\ast\check{H}}\,.  \label{6b.5ac}
\end{equation}

The functions $\psi _{\ast }(x)\in D_{\ast \check{H}}$ vanish at the origin, 
$\psi _{\ast }(x)\rightarrow 0$ as $x\rightarrow 0$, which is quite natural
from the physical standpoint for a strongly repulsive Calogero potential (%
\ref{6.0}) with $\alpha \geq 3/4\,(\varkappa >1)$. The standard heuristic
physical arguments when finding eigenfunctions in the Calogero potential are
based on a realistic hypothesis that the behavior of the eigenfunctions at
the origin is defined by homogeneous equation $\check{H}\psi _{E}=0$. Its
solution, see (\ref{6a.10}), yields $\psi _{E}(x)\simeq
c_{1}x^{1/2+\varkappa }+c_{2}x^{1/2-\varkappa }$,$\;x\rightarrow 0$. The
second term in the r.h.s. must be omitted, because it is not
square-integrable at the origin for $\varkappa >1$ unless $c_{2}=0$, and $%
\psi _{E}(x)$ must vanish as $x\rightarrow 0$. But we can represent this
vanishing at $x\rightarrow 0$ more precisely, see (\ref{7b.4a}), (\ref{7b.4b}%
),%
\begin{eqnarray}
\psi _{\ast }(x) &=&O(x^{3/2}),\ \psi _{\ast }^{\prime
}(x)=O(x^{1/2}),\;\alpha >3/4,  \notag \\
\psi _{\ast }(x) &=&O(x^{3/2}\sqrt{\left\vert \ln x\right\vert }),\ \psi
_{\ast }^{\prime }(x)=O(x^{1/2}\sqrt{\left\vert \ln x\right\vert }),\;\alpha
=3/4;  \label{6b.5a}
\end{eqnarray}%
this estimate also holds true for the eigenfunctions $\psi _{E}$.

\subsection{Second region: $-1/4<\protect\alpha <3/4\,(0<\varkappa <1)$}

In this region, the asymmetry form $\Delta _{H^{+}}$ is 
\begin{equation*}
\Delta _{H^{+}}\left( \psi _{\ast }\right) =-2k_{0}\varkappa \left( 
\overline{c_{1}}c_{2}-\overline{c_{2}}c_{1}\right) =ik_{0}\varkappa \left(
\left\vert c_{+}\right\vert ^{2}-\left\vert c_{-}\right\vert ^{2}\right)
\,,\ c_{\pm }=c_{1}\pm ic_{2};
\end{equation*}%
where $c_{\pm }$ are the diagonal a.b. coefficients\footnote{%
In passing, we find that the deficiency indices of the initial symmetric
operator $\hat{H}$ with $\alpha $ such that $-1/4<\alpha <3/4$ are $(1,1)$
and thus there must exist a one-parameter family of s.a. extensions of $\hat{%
H}$.}. In such a case, the matrix$\ U\,$\ in (\ref{6b.5d}) is reduced to a
complex number of unit module, an element of the group $U(1)$, that is a
circle, and relation (\ref{6b.5d}) becomes $c_{-}=e^{i\vartheta
}c_{+}\,,\;0\leq \vartheta \leq 2\pi \,,\;0\sim 2\pi \,,$ or, equivalently%
\footnote{%
The symbol $\overline{\mathbb{R}}$ here denotes the compactified real axis
where $-\infty $ and $+\infty $ are identified: $\overline{\mathbb{R}}%
\mathbb{=\{\lambda }:-\infty \leq \lambda \leq +\infty ,\;-\infty \sim
+\infty \mathbb{\}}$; $\overline{\mathbb{R}}$ is homeomorphic to a circle.},%
\begin{equation}
c_{2}=\lambda c_{1}\,,\;\lambda =-\tan \vartheta /2\,\in \overline{\mathbb{R}%
}\,,  \label{6b.5as}
\end{equation}%
and $\left\vert \lambda \right\vert =\infty $ means $c_{1}=0$, $c_{2}$ is
arbitrary.

The relation (\ref{6b.5as}) with any fixed $\lambda$ defines a maximum
subspace $D_{2,\lambda}\subset D_{\ast\check{H}}$ where the asymmetry form $%
\Delta_{H^{+}}$ vanishes identically. The subspace $D_{2,\lambda}$ is the
domain of an s.a. operator $\hat{H}_{2,\lambda}$, $D_{2,\lambda}=D_{H_{2,%
\lambda}}$, specified by a.b. conditions at the origin

\begin{align}
& \psi_{2,\lambda}(x)=\psi_{2,\lambda}^{\mathrm{as}}(x)+O(x^{3/2}),\ \
\psi_{2,\lambda}^{\prime}(x)=\psi_{2,\lambda}^{\mathrm{as}\prime
}(x)+O(x^{1/2})\ \mathrm{as}\;x\rightarrow0,  \notag \\
& \psi_{2,\lambda}^{\mathrm{as}}(x)=\left\{ 
\begin{array}{l}
c[(k_{0}x)^{1/2+\varkappa}+\lambda(k_{0}x)^{1/2-\varkappa}]\,,\ |\lambda
|<\infty, \\ 
c(k_{0}x)^{1/2-\varkappa}\,,\ |\lambda|=\infty~.%
\end{array}
\right. \,  \label{6b.7da}
\end{align}
\ 

We thus obtain that in the case of $-1/4<\alpha <3/4$ constructing an s.a.
Hamiltonian associated with the s.a. Calogero differential expression $%
\check{H}$ (\ref{6a.2}) is nonunique: there exists a one-parameter $U\left(
1\right) $-family $\left\{ \hat{H}_{2,\lambda },\;|\lambda |\leq \infty
\right\} $ of s.a. Hamiltonians, and their domains $D_{H_{2,\lambda }\text{ }%
}$ are given by%
\begin{equation}
D_{H_{2,\lambda }}=\left\{ \psi _{\lambda }:\psi _{\lambda }\in D_{\ast 
\check{H}};\ \psi _{\lambda },\psi _{\lambda }^{\prime }\ \mathrm{satisfy\;}%
\left( \ref{6b.7da}\right) \right\} .  \label{6b.7ad}
\end{equation}%
A concrete choice of $\lambda $, and therefore that of an s.a. Hamiltonian,
requires additional arguments.

This problem is well-known in physics, based on additional physical
considerations \cite{LanLiQM}. According to the above physical arguments, an
asymptotic behavior of eigenfunctions at the origin is given by the previous
formula $\psi _{E}(x)\simeq c_{1}x^{1/2+\varkappa }+c_{2}x^{1/2-\varkappa }$,%
$\;x\rightarrow 0$, but now $0<\varkappa <1$; both terms are therefore
square-integrable at the origin, and there arises an unexpected uncertainty
in the choice of boundary conditions at the origin and that of scattering
states. To avoid this difficulty, it is proposed to consider the regularized
cut-off potential (\ref{6a.3}) with the standard boundary conditions $\psi
_{E}(0)=0$ and the subsequent limit $r_{0}\rightarrow 0$. This limiting
procedure yields $c_{2}=0$, or the choice $\lambda =0$ in the a.b.
conditions (\ref{6b.7da}). Under this choice, we have customary zero
boundary conditions at the origin for wave functions for all $\alpha $ in
the interval $(-1/4,3/4)$, while for their derivatives we have zero boundary
conditions if $0<\alpha <3/4$ (repulsion) and specific singularities at the
origin if $-1/4<\alpha <$ $0$ (attraction). With $\lambda \neq 0$, we have
specific singularities at the origin for wave functions if $0<\alpha <3/4$
and zero boundary conditions if $-1/4<\alpha <$ $0$, while for the
derivatives of wave functions we have specific singularities for all $\alpha 
$. We make a remark on a possible physical meaning of s.a. Hamiltonians with 
$\lambda \neq 0$, which cannot be obtained by the above regularization
procedure. If we treat a one-dimensional Calogero Hamiltonian as a radial
Hamiltonian in the three- or two-dimensional Calogero problem where $x=r$,
the additional terms with the factor $\lambda $ in a.b. conditions can be
treated as a manifestation of additional singular terms of zero radius in
the potential; we call them $\delta $-like terms. There are different
arguments in favor of this suggestion. First, such potentials are not
grasped by an initial symmetric operator, whose domain is a set of functions
vanishing at the origin. Second, even in the case of free motion, $\alpha =0$%
, the a.b. conditions with $\lambda \neq 0$ are admissible, and it was first
shown in \cite{BerFa61} that a local potential $V\sim $ $\delta ^{(3)}(%
\mathbf{r})\;$can be properly treated in terms of s.a. extensions, with $%
\lambda \neq 0$, of the Laplacian $\Delta _{3}$, initially defined as a
symmetric operator on functions vanishing at the origin. An additional
heuristic argument is that three-dimensional radial functions
square-integrable with the measure $r^{2}dr$ differ from our one-dimensional
functions by the factor $1/r$; therefore if $\lambda \neq 0$ the asymptotic
behavior of the $s$-wave radial function is given by $\lambda /r$, and we
formally obtain $\Delta _{3}\lambda /r\sim $ $\lambda \delta ^{(3)}(\mathbf{r%
})$. Finally, a credible speculation is that $\delta $-like terms can be
reproduced, and therefore any s.a. Hamiltonian $\hat{H}_{2,\lambda }$ can be
obtained by means of a limiting procedure $r_{0}\rightarrow 0$ if we start
with a more sophisticated regularized potential where the cut-off potential (%
\ref{6a.3}) is supplemented by an attractive or repulsive potential of the
same radius (a square well or a core, respectively) whose strength is
appropriately fitted to $r_{0}$ in terms of a certain finite $\lambda $ that
survives in a.b. conditions in the limit $r_{0}\rightarrow 0$. A
verification of this hypothesis is an interesting problem for a further
study.

\subsection{Third region: the point $\protect\alpha =-1/4\,(\varkappa =0)$}

The consideration of this value of $\alpha $ and the result are completely
similar to those in the previous subsection .

The expression for the asymmetry form is identical to that of the previous
subsection\footnote{%
We can repeat the remark in footnote 10 for this value of $\alpha $.}:%
\begin{equation*}
\Delta _{H^{+}}=-k_{0}\left( \overline{c_{1}}c_{2}-\overline{c_{2}}%
c_{1}\right) =\frac{i}{2}k_{0}\left( \left\vert c_{+}\right\vert
^{2}-\left\vert c_{-}\right\vert ^{2}\right) \,,\;c_{\pm }=c_{1}\pm ic_{2},
\end{equation*}%
the relation between $c_{-}$ and $c_{+}$ under which $\Delta _{H^{+}}$
vanishes identically is $c_{-}=e^{i\vartheta }c_{+}\,,\;0\leq \vartheta \leq
2\pi \,,\;0\sim 2\pi \,,$ or, equivalently, 
\begin{equation}
c_{1}=\lambda c_{2}\,,\;\lambda =-\cot \vartheta /2\,,\;-\infty \leq \lambda
\leq \infty \,,\;-\infty \sim \infty \,.  \label{6b.8ac}
\end{equation}

Relation (\ref{6b.8ac}) with any fixed $\lambda $ defines a.b. conditions at
the origin,%
\begin{align}
& \psi _{3,\lambda }(x)=\psi _{3,\lambda }^{\mathrm{as}}(x)+O(x^{3/2}),\ \
\psi _{3,\lambda }^{\prime }(x)=\psi _{3,\lambda }^{\mathrm{as}\prime
}(x)+O(x^{1/2})\,\,\mathrm{as}\;x\rightarrow 0,  \notag \\
& \psi _{3,\lambda }^{\mathrm{as}}(x)=\left\{ 
\begin{array}{l}
c[\lambda x^{1/2}+x^{1/2}\ln \left( k_{0}x\right) ],\ |\lambda |<\infty , \\ 
cx^{1/2},\ |\lambda |=\infty ,%
\end{array}%
\right.  \label{6b.10}
\end{align}%
specifying an s.a. operator $\hat{H}_{3,\lambda }$.

The final conclusion is similar to that of the previous subsection: there
exists a one-parameter $U\left( 1\right) $-family $\left\{ \hat{H}%
_{3,\lambda },\;|\lambda |\leq \infty \right\} $ of s.a. Calogero
Hamiltonians associated with the s.a. Calogero differential expression $%
\check{H}$ (\ref{6a.2}) with $\alpha =$ $-1/4$, and their domains $%
D_{H_{3,\lambda }\text{ }}$are given by%
\begin{equation}
D_{H_{3,\lambda }}=\left\{ \psi _{3,\lambda }:\psi _{3,\lambda }\in D_{\ast 
\check{H}}\ ;\ \psi _{3,\lambda },\psi _{3,\lambda }^{\prime }\;\mathrm{%
satisfy\ }\left( \ref{6b.10}\right) \right\} .  \label{6b.11}
\end{equation}

We can also add a remark, similar to that in the end of the previous
subsection, on a concrete choice of$\,\lambda $ and on a possible physical
meaning of the latter. The case $\alpha =-1/4$ corresponds to a free motion
of a particle in the two-dimensional space, and the two-dimensional radial
functions square-integrable with the measure $rdr$ differ from our
one-dimensional functions by the factor $r^{-1/2}$ ; therefore, if $%
\left\vert \lambda \right\vert \neq \infty $, the asymptotic behavior of the 
$s$-wave radial function is given by $\lambda ^{-1}\ln \left( k_{0}r\right) $%
, and we formally obtain $\Delta _{2}\lambda ^{-1}\ln \left( k_{0}r\right)
\sim \lambda ^{-1}\delta ^{(2)}(\mathbf{r})$.

\subsection{Fourth region: $\protect\alpha <-1/4\;(\varkappa =i\protect%
\sigma ,\,\protect\sigma >0)$}

A consideration for these values of $\alpha $ is a copy of those in the
previous two subsections.\footnote{%
We can repeat the remark in footnote 10 for this value of $\alpha $.}

The asymmetry form is a canonical diagonal form from the very beginning:%
\begin{equation*}
\Delta _{H^{+}}\left( \psi _{\ast }\right) =i2k_{0}\sigma \left( \left\vert
c_{1}\right\vert ^{2}-\left\vert c_{2}\right\vert ^{2}\right) \,.
\end{equation*}

It follows that there exists a one-parameter $U\left( 1\right) $-family $%
\left\{ \hat{H}_{4,\theta },\;0\leq \theta \leq \pi ,\ 0\sim \pi \right\} $
of s.a. Hamiltonians associated with the s.a. Calogero differential
expression $\check{H}$ (\ref{6a.2}) with $\alpha <-1/4$ and specified by
a.b. conditions at the origin, 
\begin{align}
& \psi _{4,\theta }(x)=\psi _{\theta }^{\mathrm{as}}(x)+O(x^{3/2})\,,\ \psi
_{4,\theta }^{\prime }(x)=\ \psi _{\theta }^{\mathrm{as}\prime
}(x)+O(x^{1/2}),\mathrm{\;}x\rightarrow 0,  \notag \\
& \psi _{4,\theta }^{\mathrm{as}}(x)=cx^{1/2}\left[ e^{i\theta }\left(
k_{0}x\right) ^{i\sigma }+e^{-i\theta }\left( k_{0}x\right) ^{-i\sigma }%
\right] \,.  \label{6b.14}
\end{align}%
The domain of each of the Hamiltonian $\hat{H}_{4,\theta }$ is given by%
\begin{equation}
D_{H_{4,\theta }}=\left\{ \psi _{4,\theta }:\psi _{4,\theta }\in D_{\ast 
\check{H}}\ ;\ \psi _{4,\theta },\psi _{4,\theta }^{\prime }\;\mathrm{%
satisfy\;}\left( \ref{6b.14}\right) \right\} .  \label{6b.15}
\end{equation}%
The leading terms in a.b. conditions (\ref{6b.14}) can be written as follows:%
\begin{equation}
\psi _{4,\theta }^{\mathrm{as}}(x)=cx^{1/2}\cos (\sigma \ln k_{0}x+\theta ),
\label{6b.16}
\end{equation}%
and $2\theta $ can be interpreted as the phase of the scattered wave at the
origin. This form of a.b. conditions was first proposed in \cite{Case50}.

It is interesting to compare the result with conventional physical
considerations. The standard arguments concerning the asymptotic behavior of
eigenfunctions $\psi _{E}$ at the origin yield $\psi _{E}(x)\simeq
x^{1/2}(c_{1}x^{i\sigma }+c_{2}x^{-i\sigma })$,$\;x\rightarrow 0$. In
contrast to the previous cases, both terms are of the same infinitely
oscillating behavior, which does not allow definitely fixing $c_{1}$ and $%
c_{2}$, more precisely, fixing the ratio $c_{2}/c_{1}$, and thus
constructing scattering states.

An attempt to fix this ratio via the limiting procedure $r_{0}\rightarrow 0$
starting with the cut-off potential (\ref{6a.3}) fails: $c_{2}/c_{1}$ has no
limit as $r_{0}\rightarrow 0$, and there is no limit for eigenvalues and
eigenfunctions \cite{LanLiQM, Meetz64}. However, as stated in \cite{BawCo03}%
, superimposing the cut-off potential (\ref{6a.3}) with a square-well
attractive potential of the same radius, and thus changing the potential $%
V_{r_{0}}(x)$ to the potential $V_{s}(x)=-\alpha _{s}(r_{0})r_{0}^{-2}\theta
(r_{0}-x)-\alpha r^{-2}\theta (x-r_{0})$, where $\theta $ here stands for
the symbol of the known step function, allows obtaining a.b. conditions (\ref%
{6b.16}) in the limit of zero radius $r_{0}$ under an appropriate choice of
the coupling constant $\alpha _{s}(r_{0})$.

\section{Spectral analysis}

\subsection{Preliminary}

We now turn to a spectral analysis of Hamiltonians comprising the above four
families in accordance with different values of the coupling constant $%
\alpha $. This includes finding the spectrum and (generalized)
eigenfunctions for each Hamiltonian and deriving formulas for the respective
eigenfunction expansions of arbitrary square-integrable function. The short
name for these formulas in mathematics is \textquotedblleft inversion
formulas\textquotedblright .

In what follows, we use this term. In solving the spectral problem, we
follow Krein's method of guiding functionals where the spectrum and
eigenfunctions emerge in the process of deriving inversion formulas \cite%
{AkhGl81,Naima69}. For differential operators of second order, we generally
need two guiding functionals. But in the case where the spectrum is expected
to be simple, as in our case, it suffices to have only one, the so-called
simple guiding functional. We remind the reader of the basics of Krein's
method in this case\footnote{%
These were outlined in \cite{VorGiTC07}.} as applied to our problem.

Let $\hat{H}$ be an s.a. Hamiltonian associated with the differential
expression $\check{H}$ (\ref{6a.2}); by $\hat{H}$, we imply any operator in
the above four families. Krein's method for a spectral analysis of $\hat{H}$
rests on\textrm{\ }using certain solutions of the homogeneous differential
equation 
\begin{equation}
(\check{H}-W)u(x;W)=0,  \label{7c.1}
\end{equation}%
where $W=\func{Re}W+\func{Im}W=E+\func{Im}W$ is an arbitrary complex number;
we can say that $u(x;W)$ is \textquotedblleft an eigenfunction of $\check{H}$
with complex energy $W$\textquotedblright .

Let $u(x;W)$ be a function with the following properties:

i) $u(x;W)\ $is a solution of the homogeneous equation (\ref{7c.1}),

ii) $u(x;W)$ is real-entire in $W$, i.e., it is an entire function of $W$
for each fixed\textrm{\ }$x$ that is real for a real $W$: $u(x;E)\,=%
\overline{u(x;E)}$,

iii) $u(x;W)$ satisfies a.b. conditions specifying the Hamiltonian $\hat{H}$
under consideration.

Such a function certainly does exist (see below).

Let $\mathbb{D}$ be a space of functions $\xi $ belonging to the domain of $%
\hat{H}\;$and vanishing for $x>b>0$, where $b$ may be different for each $%
\xi $, i.e., $\mathbb{D=}D_{H}\cap D_{r}(\mathbb{R}_{+})$, where $D_{r}(%
\mathbb{R}_{+})$ is the space of functions in $\mathbb{R}_{+}$ with a
support bounded from the right. The space $\mathbb{D}$ is dense in $%
L^{2}(R_{+})$,\ $\overline{\mathbb{D}}\,=L^{2}(R_{+})$. The linear
functional $\Phi (\xi ;W)$ defined in a space $\mathbb{D}$ and given by%
\begin{equation}
\Phi (\xi ;W)=\int_{0}^{\infty }u(y;W)\xi (y)dy\,,\;\forall \xi \in \mathbb{D%
}\,,\text{\thinspace }  \tag{.}  \label{7c.2}
\end{equation}%
is called a guiding functional; the integration in the r.h.s. of (\ref{7c.2}%
) actually goes up to some finite $b$.

It is evident that $\Phi (\xi ;W)$ is an entire function of $W$ (in fact,
for the function $u(x;W)\,\ $to be real-entire, and thus for the functional $%
\Phi (\xi ;W)$ to be entire in $W$, it suffices to establish that $u(x;W)$
is analytic in some strip containing the real axis and that $u(x;E)$ is
real) for each fixed $\xi $ and obeys the property $\Phi (\check{H}\xi
;W)=W\Phi (\xi ;W)$, which follows from the Lagrange identity for the
functions $u$ and $\xi $ satisfying the same a.b. conditions at the origin.
Let the functional $\Phi $ also obey the property 
\begin{equation}
\Phi (\xi _{0};E_{0})=0,\;\xi _{0}\in \mathbb{D}_{H}\Longrightarrow \exists
\psi _{0}\in \mathbb{D}_{H}\emph{\,,\ }\,(\check{H}-W)\psi _{0}=\xi _{0},
\label{7c.3}
\end{equation}%
then we call $\Phi $ a simple guiding functional.

Let the functional $\Phi $ be simple. Then there hold the following
statements.

1) The spectrum of the Hamiltonian $\hat{H}$, \textrm{spec} $\hat{H}$, is
simple.

2) There exists a spectral function $\sigma(E)$ for the Hamiltonian $\hat{H}$
such that \textrm{spec} $\hat{H}$ is the set of its growth points. The
spectral function defines the Hilbert space $L_{\sigma}^{2}\;$with the
measure $d\sigma(E)$.

3) For any function $\psi \in L^{2}(\mathbb{R}_{+})$, the inversion formulas 
\begin{equation}
\psi (x)=\int \varphi (E)u\left( x;E\right) d\sigma (E)\,,\ \varphi
(E)=\int_{0}^{\infty }u\left( x;E\right) \psi (x)dx\,\in L_{\sigma }^{2}
\label{7c.4b}
\end{equation}%
hold true, together with the Parseval equality%
\begin{equation}
\int_{0}^{\infty }|\psi (x)|^{2}dx=\int |\varphi (E)|^{2}d\sigma (E)\,.
\label{7c.4c}
\end{equation}

The integrals in the r.h.s. of (\ref{7c.4b}) converge in the respective $%
L^{2}(\mathbb{R}_{+})$ and $L_{\sigma}^{2}$. The integration over $E$ in the
r.h.s. of (\ref{7c.4b}), (\ref{7c.4c}) goes over \textrm{\ spec }$\hat{H}$,
so that we can set $\varphi(E)=0$ if $E\notin$\textrm{\ spec }$\hat{H}$,
i.e., for all constancy points of $\sigma(E)$, and $u\left( x;E\right) $ for
such $E$ do not enter the inversion formulas.

4) Let the spectral function be the sum of a jump function $\sigma _{jmp}(E)$
and of an absolutely continuous function $\sigma _{a.c.}(E)$, $\sigma
(E)=\sigma _{jmp}(E)+\sigma _{a.c.}(E)$, as in our case\footnote{%
The spectral function does not contain so-called singular terms.}. Then $%
d\sigma (E)$ can be represented as $d\sigma (E)=\sigma ^{\prime }(E)dE$,
where the derivative $\sigma ^{\prime }(E)$, the so-called spectral density,
is understood in the distribution sense and is given by%
\begin{equation*}
\sigma ^{\prime }(E)=\sum_{n}\rho _{n}\,\delta (E-E_{n})+\rho _{c}(E),\;\rho
_{n}>0,\mathrm{\,}\rho _{c}(E)\geq 0.\,
\end{equation*}

The set $\left\{ E_{n}\right\} $ can be empty; if not, the real numbers $%
E_{n},\,\left\{ n\right\} \subset \mathbb{Z}$, are the energy eigenvalues
for the Hamiltonian $\hat{H}$ corresponding to bound states. The set $%
\left\{ E_{n}\right\} $ of bound-state energies is the discrete spectrum (or
the discrete part of the spectrum) of $\hat{H}$, while the set $\mathrm{%
supp\,}\rho _{c}(E)$, is the continuous spectrum (or the continuous part of
the spectrum) of $\hat{H}$, and the whole spectrum of $\hat{H}$ is the union
of these sets $\mathrm{spec\ }\hat{H}\,=\left\{ E_{n}\right\} \cup \mathrm{%
supp\,}\rho _{c}(E).$ Accordingly, the functions $u(x;E_{n})$ are
normalizable eigenfunctions of bound states of $\hat{H}$, while $%
u(x;E),\,E\in \mathrm{supp\,}\rho _{c}(E)$, are (generalized) eigenfunctions
of the continuous spectrum of $\hat{H}$.

If we introduce normalized eigenfunctions $u_{\mathrm{nr}}(x;E)\,$by%
\begin{equation*}
u_{\mathrm{nr}}(x)=\left\{ 
\begin{array}{l}
u_{n}(x)=\sqrt{\rho _{n}}u(x;E_{n}),\,E=E_{n}, \\ 
u_{E}(x)=\sqrt{\rho (E)}u(x;E),\,E\in \mathrm{supp\,}\rho _{c}(E),%
\end{array}%
\right.
\end{equation*}%
then the inversion formulas (\ref{7c.4b}) and the Parseval equality (\ref%
{7c.4c}) become%
\begin{align}
& \psi (x)=\sum_{n}\varphi _{n}u_{n}(x)+\int_{\mathrm{supp\,}\rho
_{c}(E)}\varphi (E)u_{E}(x)dE\,,  \label{7c.8a} \\
& \varphi (E)=\int_{0}^{\infty }u_{E}\left( x\right) \psi (x)dx\,,\varphi
_{n}=u_{n}(x)\int_{0}^{\infty }u_{n}(x)\psi (x)dx,  \label{7c.8b} \\
& \int_{0}^{\infty }|\psi (x)|^{2}dx=\sum_{n}\left\vert \varphi
_{n})\right\vert ^{2}+\int_{\mathrm{supp\,}\rho _{c}(E)}\,\left\vert \varphi
(E)\right\vert ^{2}dE.\,  \label{7c.8c}
\end{align}

We note that the inversion formulas and the Parseval equality do not change
(are invariant) under the change of the sign of any of the bound-state
eigenfunctions, being the muliplication by $-1$ (and of the
continuous-spectrum eigenfunctions), which can sometimes be convenient.

In what follows, when completing the spectral analysis of each family of
s.a. Calogero Hamiltonians, we do not present the respective inversion
formulas explicitly, but restrict ourselves to an assertion of the type
\textquotedblleft the above-given set of eigenfunctions form a complete
orthonormalized system of eigenfunctions for the Hamiltonian under
consideration\textquotedblright ; by such an assertion we mean that there
hold inversion formulas and a Parseval equality of the form (\ref{7c.8a}), (%
\ref{7c.8b}), and (\ref{7c.8c}), in terms of the corresponding normalized
eigenfunctions. Using this terminology, we follow the physical tradition.
Formulas (\ref{7c.8a})--(\ref{7c.8c}) are of customary form for physicists.
From the physical standpoint, these formulas testify that the eigenfunctions
form a complete orthogonal system in the sense that they satisfy the
respective completeness and orthonormality relations 
\begin{align}
& \sum_{n}u_{n}\left( x\right) u_{n}(x^{\prime })+\int_{\mathrm{c.spec}\,%
\hat{H}}u_{E}\left( x\right) u_{E}(x^{\prime })dE=\delta (x-x^{\prime }),
\label{7c.9a} \\
& \int_{0}^{\infty }u_{n}\left( x\right) u_{n^{\prime }}\left( x\right)
dx=\delta _{nn^{\prime }},\ \int_{0}^{\infty }u_{n}\left( x\right)
u_{E}\left( x\right) dx=0,  \notag \\
& \int_{0}^{\infty }u_{E}\left( x\right) u_{E^{\prime }}\,\left( x\right)
dx=\delta (E-E^{\prime }),\ \,E,E^{\prime }\in \mathrm{c.spec}\,\hat{H},
\label{7c.9b}
\end{align}%
where $\mathrm{c.spec}\,\hat{H}\mathrm{\ }$denote the continuous spectrum of 
$\hat{H}$; here, $\ \mathrm{c.spec}\,\hat{H}=\mathrm{supp\,}\rho _{c}(E)$.
In physical texts on QM, the main effort is usually made to establish
precisely these relations, and the last relation in (\ref{7c.9b}) is
conventionally called \textquotedblleft the normalization of the
continuous-spectrum eigenfunctions to the $\delta $ function%
\textquotedblright . It is needless to say that, from the mathematical
standpoint, the relations in (\ref{7c.9a}) and (\ref{7c.9b}) containing $%
\delta $ functions are at most heuristic.

5) The spectral function $\sigma (E)$ is evaluated via Green's function $%
G(x,y;W)$ of the Hamiltonian $\hat{H}$ that is the integral kernel of the
resolvent $\hat{R}(W)=(\hat{H}-W)^{-1}$, i.e., a kernel of the integral
representation $\psi (x)=\int_{0}^{\infty }G(x,y;W)\chi (y)dy$ for a unique
solution $\psi \in D_{H}$ of the nonhomogeneous equation $(\check{H}-W)\psi
=\chi $ with an arbitrary square-integrable r.h.s. $\chi ,\,\chi \in L^{2}(%
\mathbb{R}_{+})$. It suffices to consider $W$ in the upper half-plane, $%
\func{Im}W>0$. Namely, if we introduce the function 
\begin{equation}
M(c;W)=G(c-0,c+0;W),  \label{7c.11a}
\end{equation}%
where $c$ is an arbitrary inner point of $\mathbb{R}_{+}$, $c\in (0,\infty )$%
, then the spectral density $\sigma ^{\prime }(E)$ is determined by the
relation 
\begin{equation}
\left[ u(c;E)\right] ^{2}\sigma ^{\prime }(E)=\pi ^{-1}\func{Im}M(c;E+i0).
\label{7c.12}
\end{equation}

It remains to find a convenient representation for Green's function that
allows finding the spectral density.

Let $v(x;W)$ be a solution of the homogeneous equation (\ref{7c.1}) that is
linearly independent of the solution $u(x;W)$ and exponentially decreasing
at infinity (it certainly does exist). Then Green's function is given by%
\begin{equation}
G(x,y;W)=\omega ^{-1}(W)\left\{ 
\begin{array}{l}
v(x;W)\,u(y;W),\;x>y, \\ 
u\left( x;W\right) \,v(y;W),\;x<y,%
\end{array}%
\right.  \label{7c.13}
\end{equation}%
where $\omega (W)=-\mathrm{Wr}(u,v),$ and therefore, the function $M(c;W)\,$(%
\ref{7c.11a}) is given by%
\begin{equation}
M(c;W)=\omega ^{-1}(W)u\left( c;W\right) \,v(c;W).  \label{7c.14a}
\end{equation}

Formulas (\ref{7c.12}), (\ref{7c.13}) and (\ref{7c.14a}) suffice for
evaluating the spectral function. However, from the calculation standpoint,
the following modification may appear to be more suitable. Let $\widetilde{u}%
(x;W)$ be a solution of the homogeneous equation (\ref{7c.1}) that is
real-entire, as well as the solution $u(x;W)$, but is linearly independent
of $u(x;W)$, so that their Wronskian is $\mathrm{Wr}(u,\widetilde{u})=-%
\widetilde{\omega }(W)\neq 0$, and the function $\widetilde{\omega }(W)$ is
real-entire. The functions $u$ and $\widetilde{u}$ form a fundamental set of
solutions of the homogeneous equation (\ref{7c.1}); therefore, the function $%
v$ allows the representation 
\begin{align}
& v(x,W)=c_{1}(W)u(x;W)+c_{2}(W)\widetilde{u}(x;W),\,  \notag \\
& c_{1}(W)=-\frac{\mathrm{Wr}(v,\widetilde{u})}{\widetilde{\omega }(W)}%
,\,c_{2}(W)=\frac{\omega (W)}{\widetilde{\omega }(W)},  \label{7c.15} \\
& c_{1,2}(W)\neq 0\ \mathrm{for}\ \func{Im}W>0.  \notag
\end{align}%
We note that the function $v$ is defined up to a nonzero factor, and we can
make a change $v\rightarrow v/c_{1}$; then representation (\ref{7c.15})
becomes%
\begin{equation}
v(x,W)=u(x;W)+\frac{\omega (W)}{\widetilde{\omega }(W)}\widetilde{u}(x;W).
\label{7c.16}
\end{equation}%
With such a choice of $v$ and (\ref{7c.13}), (\ref{7c.14a}) taken into
account, Eq. (\ref{7c.12}) becomes%
\begin{equation}
\sigma ^{\prime }(E)=\pi ^{-1}\lim_{\varepsilon \rightarrow +0}\func{Im}%
\omega ^{-1}(E+i\varepsilon )  \label{7c.17}
\end{equation}%
because $u(c;E+i0)=$ $u(c;E)$,\thinspace\ $\widetilde{u}(c;E+i0)=\widetilde{u%
}(c;E)$, and $\widetilde{\omega }(E+i0)=\widetilde{\omega }(E)$ are real.

We should clarify formula (\ref{7c.17}). The function $\omega (E)=\omega
(E+i0)=\lim_{\varepsilon \rightarrow +0}\omega (E+i\varepsilon )$ can have
simple isolated zeroes. If $\omega (E)\neq 0$, then $\lim_{\varepsilon
\rightarrow +0}\func{Im}\omega ^{-1}(E+i\varepsilon )=\func{Im}\omega
^{-1}(E)$, and therefore,%
\begin{equation}
\sigma ^{\prime }(E)=\pi ^{-1}\lim_{\varepsilon \rightarrow +0}\func{Im}%
\omega ^{-1}(E)\,\mathrm{\,if\,\,}\omega (E)\neq 0.  \label{7c.17b}
\end{equation}%
Let $E_{0}$ be a simple isolated zero of the function $\omega (E),$ $\omega
(E_{0})=0$. The function $\omega ^{-1}(W)$ allows the representation%
\begin{equation*}
\omega ^{-1}(W)=\left[ \omega ^{\prime }(E_{0})\left( W-E_{0}\right) \right]
^{-1}+\phi (W),\;\omega ^{\prime }(E_{0})<0,
\end{equation*}%
where the function $\phi (W)=\omega ^{-1}(W)-\left[ \omega ^{\prime
}(E_{0})\left( W-E_{0}\right) \right] ^{-1}$ is nonsingular in some
neighborhood of the point $E_{0}$. Using the known formula%
\begin{equation*}
\lim_{\varepsilon \rightarrow +0}\func{Im}\left( E-E_{0}+i\varepsilon
\right) ^{-1}=-\pi \delta (E-E_{0}),
\end{equation*}%
we then obtain that in this neighbourhood there holds the representation 
\begin{equation*}
\lim_{\varepsilon \rightarrow +0}\func{Im}\omega ^{-1}(E+i\varepsilon )=-%
\frac{\pi }{\omega ^{\prime }(E_{0})}\delta (E-E_{0})+\func{Im}\phi (E),
\end{equation*}%
where%
\begin{equation*}
\func{Im}\phi (E)=\func{Im}\omega ^{-1}(E),\ E\neq E_{0},\ \mathrm{\,}\func{%
Im}\phi (E_{0})=\lim_{E\rightarrow E_{0}}\func{Im}\phi (E),
\end{equation*}%
in particular, $\func{Im}\phi (E)=0$ if $\omega (E)$ is real, and we find
that%
\begin{equation}
\sigma ^{\prime }(E)=-(\omega ^{\prime }(E_{0}))^{-1}\delta (E-E_{0})\,\,%
\mathrm{if\,\,}\omega (E_{0})=0\,,  \label{7c.17c}
\end{equation}%
and$\mathrm{\ \func{Im}}\omega (E)=0\mathrm{\,}$in a neighbourhood of $%
E_{0}. $ After this, we proceed with a direct spectral analysis of the s.a.
Calogero Hamiltonians, sequentially from the first family to the fourth one
in accordance with the different regions of values of the coupling constant $%
\alpha $. In each region, the spectral analysis and its result have some
specific features. We also remember that in each case we must verify that a
chosen guiding functional is simple, i.e., that property (\ref{7c.3}) for a
given guiding functional holds true. Because a simple substitution reduces
the homogeneous equation (\ref{7c.1}) to the Bessel equation, see (\ref{6a.8}%
), (\ref{6a.11}) and (\ref{6a.12}), the above functions $u,v$ and $%
\widetilde{u}$ are different Bessel functions up to the factor $x^{1/2}$; we
hope that the cited properties of Bessel functions are well-known or can be
easily taken out of handbooks on special functions. The necessary Wronskians
can be evaluated using the asymptotic expansions of the corresponding
functions at the origin. As a rule, we label all the functions involved by
indices indicating a family and an extension parameter, as well as the
corresponding Hamiltonians.

\subsection{First region: $\protect\alpha \geq 3/4\,(\varkappa \geq 1)$}

In this region, there exists only one s.a. Calogero Hamiltonian $\hat{H}_{1}$
defined on the natural domain $D_{H_{1}}=D_{\ast\check{H}}$.

For the functions $u$ and $v$, the above-described special solutions of the
homogeneous equation (\ref{7c.1}) are taken as the respective 
\begin{align}
u_{1}\left( x;W\right) & =\left( \beta /2k_{0}\right) ^{-\varkappa
}x^{1/2}J_{\varkappa }(\beta x),  \label{7c.18a} \\
v_{1}\left( x;W\right) & =\left( \beta /2k_{0}\right) ^{\varkappa
}x^{1/2}H_{\varkappa }^{(1)}(\beta x),  \label{7c.18b}
\end{align}%
where $J_{\varkappa }$ is the Bessel function; $H_{\varkappa }^{(1)}$ is the
Hankel function, and we set $W=\left\vert W\right\vert \exp i\varphi $,
\thinspace $0<\varphi <\pi $, $\,\beta =\sqrt{W}=\sqrt{\left\vert
W\right\vert }\exp i\varphi /2$, $\func{Im}\beta >0$, while $k_{0}$ is a
(fixed) parameter of a dimension of the inverse length introduced by
dimensional reasons. The asymptotic behavior of these functions at the
origin, as $x\rightarrow 0$, is given by%
\begin{align}
u_{1}\left( x;W\right) & =\frac{k_{0}^{-1/2}}{\Gamma (1+\varkappa )}%
(k_{0}x)^{1/2+\varkappa }[1+O(x^{2})],  \label{7c.19a} \\
v_{1}\left( x;W\right) & =-i\frac{k_{0}^{-1/2}\Gamma (\varkappa )}{\pi }%
(k_{0}x)^{1/2-\varkappa }[1+O(x^{2})],  \label{7c.19b}
\end{align}%
whence it follows that $\omega _{1}(W)=-\mathrm{Wr}(u_{1},v_{1})=-2i/\pi .$
Their asymptotic behavior at infinity, as $x\rightarrow \infty $ , is given
by%
\begin{align}
u_{1}\left( x;W\right) & =\frac{1}{2\sqrt{\pi k_{0}}}\left( 2k_{0}/\beta
\right) ^{1/2+\varkappa }e^{-i(\beta x-\varkappa \pi /2-\pi
/4)}[1+O(x^{-1)}]\rightarrow \infty ,  \notag \\
v_{1}\left( x;W\right) & =\frac{2\sqrt{k_{0}}}{\beta \sqrt{\pi }}\left(
\beta /2k_{0}\right) ^{1/2+\varkappa }e^{i(\beta x-\varkappa \pi /2-\pi
/4)}[1+O(x^{-1)}]\rightarrow 0.  \label{7c.20b}
\end{align}

It is easy to see from (\ref{7c.18a}) and (\ref{7c.19a}) that the function $%
u_{1}\left( x;W\right) $ is real-entire in $W$ and obeys the required a.b.
conditions (\ref{6b.5a}). The guiding functional is%
\begin{equation}
\Phi (\xi ;W)=\int_{0}^{\infty }u_{1}(y;W)\xi (y)dy\,,\;\xi \in \mathbb{D}%
=D_{\ast \check{H}}\left( \mathbb{R}_{+}\right) \cap D_{r}(\mathbb{R}_{+}).
\label{7c.21}
\end{equation}%
We check that $\Phi $ meets property (\ref{7c.3}). Let%
\begin{equation}
\Phi (\xi _{0};E_{0})=0,\;\xi _{0}\in \mathbb{D}\,.  \label{7c.22}
\end{equation}%
Because $\xi _{0}\in \mathbb{D}$, its support is bounded, \textrm{supp }$\xi
_{0}\subseteq \lbrack 0,b]$ with some $b<\infty $, and (\ref{7c.22}) is
equivalent to%
\begin{equation}
\int_{0}^{b}u_{1}(y;W)\xi _{0}(y)dy=0.  \label{7c.23}
\end{equation}%
We consider the function $\psi _{0}$ given by%
\begin{equation}
\psi _{0}(x)=\frac{i\pi }{2}\left[ v_{1}(x;E_{0})\int_{0}^{x}u_{1}(y;E_{0})%
\xi _{0}(y)dy+u_{1}(x;E_{0})\int_{x}^{b}v_{1}(y;E_{0})\xi _{0}(y)dy\right] .
\label{7c.24a}
\end{equation}%
This function evidently satisfies the equation $(\check{H}-E_{0})$ $\psi
_{0}=\xi _{0}$, and%
\begin{equation}
\psi _{0}^{\prime }(x)=\frac{i\pi }{2}\left[ v_{1}^{\prime
}(x;E_{0})\int_{0}^{x}u_{1}(y;E_{0})\xi _{0}(y)dy+u_{1}^{\prime
}(x;E_{0})\int_{x}^{b}v_{1}(y;E_{0})\xi _{0}(y)dy\right] .  \label{7c.24b}
\end{equation}%
If we take account of ( \ref{7c.19a}) and (\ref{7c.19b}), the asymptotic
behavior of $\psi _{0}$ and $\psi _{0}^{\prime }$ at the origin is simply
estimated by means of the Cauchy--Bunyakovskii inequality for the integrals
in (\ref{7c.24a}) and (\ref{7c.24b}), to yield $\psi (x)=O(x^{3/2})$ and $%
\psi ^{\prime }(x)=O(x^{1/2})$ as $x\rightarrow 0$.

On the other hand, for sufficiently large $x$ we have%
\begin{equation*}
\psi _{0}(x)=\frac{i\pi }{2}v_{1}(x;E_{0})\int_{0}^{b}u_{1}(y;E_{0})\xi
_{0}(y)dy=0,\,\ x>b,
\end{equation*}%
in view of (\ref{7c.23}), i.e., \textrm{supp }$\psi _{0}\subseteq \lbrack
0,b]$. This means that $\psi _{0}\in \mathbb{D}$ and the guiding functional\
(\ref{7c.21}) is therefore simple.

The function $M$ (\ref{7c.14a}) in this region is 
\begin{equation*}
M(c;W)=\frac{i\pi }{2}u_{1}\left( c;W\right) v_{1}\left( c;W\right) =\frac{%
i\pi }{2}cJ_{\varkappa }(\beta c)H_{\varkappa }^{(1)}(\beta c)
\end{equation*}

Let $E=-\tau ^{2}<0$, $\tau >0,\ \beta =e^{i\pi /2}\tau $. Using the
representations%
\begin{equation}
J_{\varkappa }((e^{i\pi /2}\tau x)=e^{\frac{i\varkappa \pi }{2}}I_{\varkappa
}(\tau x),\;H_{\varkappa }^{(1)}(e^{i\pi /2}\tau x)=e^{-\frac{i\varkappa \pi 
}{2}}\frac{2}{i\pi }K_{\varkappa }(\tau x)\,,  \label{7c.25ab}
\end{equation}%
where $I_{\varkappa }$ is the modified Bessel function (of the first kind)
and $K_{\varkappa }$ is the McDonald function\footnote{%
Other names are the Bessel functions of imaginary arguments.}, which are
real for real arguments, we find $\func{Im}M(c;E+i0)=0$,\ $\,E<0.$

Let $E=p^{2}\geq 0$, $\beta =\sqrt{E}=p\geq 0$. Using the representation $%
H_{\varkappa }^{(1)}(px)=J_{\varkappa }(px)+iN_{\varkappa }(px)$, where $%
N_{\varkappa }$ is the Neumann function, which is real for real arguments,
we find%
\begin{equation}
\func{Im}M(c;E+i0)=\frac{\pi }{2}cJ_{\varkappa }^{2}(\sqrt{E}c)=\frac{\pi }{2%
}\left( E/4k_{0}^{2}\right) ^{\varkappa }\left[ u_{1}(c;E)\right] ^{2},\
\,E\geq 0.  \label{7c.25a}
\end{equation}%
Using (\ref{7c.25a}), we obtain $\sigma ^{\prime }(E)=2^{-1}\left(
E/4k_{0}^{2}\right) ^{\varkappa },\ E\geq 0.$

This means that the energy spectrum of the s.a. Calogero Hamiltonian $\hat{H}%
_{1}$ is the semiaxis $\mathbb{R}_{+}$, $\mathrm{spec\,}\hat{H}%
_{1}=[0,\infty ),$ continuous and simple.

The normalized generalized eigenfunctions%
\begin{equation}
u_{1,E}\,\left( x\right) =\sqrt{\rho _{1}(E)}u_{1}(x;E)=\frac{1}{\sqrt{2}}%
x^{1/2}J_{\varkappa }(\sqrt{E}x),\,E\ \geq 0\ ,  \label{7c.26b}
\end{equation}%
form a complete orthonormalized system of eigenfunctions for the the s.a.
Calogero Hamiltonian $\hat{H}_{1}$.

We note that the inversion formulas in this case coincide with the known
formulas for the Fourie--Bessel transformation; see, for example, \cite%
{DitPr61,Akhie63}.

\subsection{Second region: $-1/4<\protect\alpha <3/4\,(0<\varkappa <1)$}

For each $\alpha $ in this region, there exists a one-parameter $U\left(
1\right) $-family of s.a. Calogero Hamiltonians $\hat{H}_{2,\lambda
},\;|\lambda |\leq \infty ,$ defined on the domains $D_{H_{2,\lambda }}$
given by (\ref{6b.7ad}).

We first note that the function 
\begin{equation}
u_{2}(x;W)=\left( \beta /2k_{0}\right) ^{\varkappa }x^{1/2}J_{-\varkappa
}(\beta x)  \label{7c.2.1}
\end{equation}%
is a solution of the homogeneous equation (\ref{7c.1}), linearly independent
of the solution $u_{1}$ (\ref{7c.18a}) (of course, with the new value of $%
\varkappa $) and is real-entire in $W$. Its asymptotic behavior at the
origin, as $x\rightarrow 0$, is given by%
\begin{equation}
u_{2}(x;W)=\frac{k_{0}^{-1/2}}{\Gamma (1-\varkappa )}(k_{0}x)^{1/2-\varkappa
}[1+O(x^{2})],  \label{7c.2.2}
\end{equation}%
and, therefore, taking (\ref{7c.19a}) and the relation $\Gamma (1+\varkappa
)\Gamma (1-\varkappa )=\pi \varkappa /\sin \pi \varkappa $ into account, we
have $\mathrm{Wr}(u_{1},u_{2})=-2\pi ^{-1}\sin \pi \varkappa .$

According to the a.b. conditions (\ref{6b.7da}), we have to distinguish the
cases of $\left\vert \lambda \right\vert <\infty $ and $\left\vert \lambda
\right\vert =\infty $. We first consider the case of $\left\vert \lambda
\right\vert <\infty $.

The a.b. conditions (\ref{6b.7da}) with $\left\vert \lambda \right\vert
<\infty $, formulas (\ref{7c.19a}) and (\ref{7c.2.2}), on the one hand, and
formulas (\ref{7c.18b}) and (\ref{7c.20b}), representation (\ref{7c.16}),
and the relation 
\begin{equation}
H_{\varkappa }^{(1)}(z)=\frac{1}{i\sin \pi \varkappa }\left[ J_{-\varkappa
}(z)-e^{-i\pi \varkappa }J_{\varkappa }(z)\right] ~,  \label{7c.2.4}
\end{equation}%
on the other hand, define the following choice for the functions $u,$ $%
\widetilde{u}$, and $v$ (see Preliminary) in the case under consideration: 
\begin{align}
& u_{2,\lambda }(x;W)=u_{1}(x,W)+\tilde{\lambda}u_{2}(x,W)  \notag \\
& =\left( \beta /2k_{0}\right) ^{-\varkappa }x^{1/2}J_{\varkappa }(\beta x)+%
\tilde{\lambda}\left( \beta /2k_{0}\right) ^{\varkappa }x^{1/2}J_{-\varkappa
}(\beta x),  \label{7c.2.5} \\
& \widetilde{u}_{2}(x;W)=u_{2}(x,W)=\left( \beta /2k_{0}\right) ^{\varkappa
}x^{1/2}J_{-\varkappa }(\beta x),  \notag \\
& v_{2}\left( x;W\right) =\frac{\sin \pi \varkappa }{i}e^{+i\pi \varkappa
}\left( \beta /2k_{0}\right) ^{-\varkappa }x^{1/2}H_{\varkappa }^{(1)}(\beta
x)  \notag \\
& =u_{1}(x,W)-\left( e^{-i\pi /2}\beta /2k_{0}\right) ^{-2\varkappa
}u_{2}(x,W)  \notag \\
& =u_{2,\lambda }(x,W)-\left[ \tilde{\lambda}+\left( e^{-i\pi /2}\beta
/2k_{0}\right) ^{-2\varkappa }\right] \widetilde{u}_{2}(x,W),\ \tilde{\lambda%
}=\frac{\Gamma (1-\varkappa )}{\Gamma (1+\varkappa )}\lambda .
\label{7c.2.6}
\end{align}%
We note that \textrm{sign }$\tilde{\lambda}=$\textrm{sign }$\lambda $.

It is easy to see that the function $u_{2,\lambda }(x;W)$ is real-entire in $%
W$ and its asymptotic behavior at the origin is given by 
\begin{equation*}
u_{2,\lambda }(x;W)=\frac{k_{0}^{-1/2}}{\Gamma (1+\varkappa )}\left[
(k_{0}x)^{1/2+\varkappa }+\lambda (k_{0}x)^{1/2-\varkappa }\right]
[1+O(x^{2})],\,x\rightarrow 0
\end{equation*}%
(which agrees with the required a.b. conditions) and that $\widetilde{\omega 
}_{2}=-\mathrm{Wr}(u_{2,\lambda },\widetilde{u}_{2})=2\pi ^{-1}\sin \pi
\varkappa $ and%
\begin{align}
\omega _{2,\lambda }(W)& =-\mathrm{Wr}(u_{2,\lambda },v_{2})=-\left[ \tilde{%
\lambda}+\left( e^{-i\pi /2}\beta /2k_{0}\right) ^{-2\varkappa }\right] 
\widetilde{\omega }_{2}  \notag \\
& =-\frac{2\sin \pi \varkappa }{\pi }\left[ \tilde{\lambda}+\left( e^{-i\pi
/2}\beta /2k_{0}\right) ^{-2\varkappa }\right] .  \label{7c.2.8}
\end{align}%
In addition, the last equality in (\ref{7c.2.6}) is a copy of the required
representation (\ref{7c.16}) with the evident substitutions $v\rightarrow
v_{2}$,\ $\,u\rightarrow u_{2,\lambda }$, $\,\omega \rightarrow \,\omega
_{2,\lambda }$, $\,\widetilde{\omega }\rightarrow \widetilde{\omega }_{2}$, $%
\,\widetilde{u}\rightarrow \widetilde{u}_{2}$.

The guiding functional is given by 
\begin{equation*}
\Phi (\xi ;W)=\int_{0}^{\infty }u_{2,\lambda }(y;W)\xi (y)dy,\ \xi \in
D_{H_{2,\lambda }}\cap D_{r}(\mathbb{R}_{+}).
\end{equation*}%
The proof of the simplicity of this guiding functional is completely similar
to that in the previous subsec.~4.2 for the first region of values of $%
\alpha $ with the replacements\ $u_{1}(x,W)\rightarrow u_{2,\lambda }(x;W)$
and $v_{1}\left( x;W\right) \rightarrow v_{2}\left( x;W\right) $.

It follows that a copy of representation (\ref{7c.17}) holds true for the
spectral density, $\sigma ^{\prime }(E)=\pi ^{-1}\lim_{\varepsilon
\rightarrow +0}\func{Im}\omega _{2,\lambda }^{-1}(E+i\varepsilon ).$

The function $\omega _{2,\lambda }(E)=\omega _{2,\lambda }(E+i0)$ is given
by 
\begin{equation}
\omega _{2,\lambda }(E)=-\frac{2\sin \pi \varkappa }{\pi }\left\{ 
\begin{array}{l}
\left[ \tilde{\lambda}+\left( -E/4k_{0}^{2}\right) ^{-\varkappa }\right]
,\;E<0,\,\, \\ 
\left( E/4k_{0}^{2}\right) ^{-\varkappa }\left[ \cos \pi \varkappa +\tilde{%
\lambda}\left( E/4k_{0}^{2}\right) ^{\varkappa }+i\sin \pi \varkappa \right]
,\,E\geq 0,%
\end{array}%
\right.  \label{D}
\end{equation}%
which shows that the function $\omega _{2,\lambda }(E)$ is real for $E<0$
and has only one negative simple zero if $\lambda <0$, while for $E\geq 0$,
it is nonzero and complex-valued. We therefore can use formulas (\ref{7c.17b}%
) and (\ref{7c.17c}) with $\omega (E)=$ $\omega _{2,\lambda }(E)$ for
evaluating the spectral density $\sigma ^{\prime }(E)$, but have to
distinguish two regions of the values of the extension parameter $\lambda $: 
$\lambda \geq 0$ and $\lambda <0$.

1. Let $\lambda \geq 0$. We then find 
\begin{equation}
\sigma ^{\prime }(E)=\frac{\theta (E)}{2\zeta _{2,\lambda }(E)}\left(
E/4k_{0}^{2}\right) ^{\varkappa },\,\ \lambda \geq 0,\ \theta (E)=\left\{ 
\begin{array}{c}
1,\ E\geq 0, \\ 
0,\ E<0,%
\end{array}%
\right.  \label{7c.2.12a}
\end{equation}%
where\textrm{\ }%
\begin{equation}
\zeta _{2,\lambda }(E)=1+2\tilde{\lambda}(E/4k_{0}^{2})^{\varkappa }\cos \pi
\varkappa +\tilde{\lambda}^{2}(E/4k_{0}^{2})^{2\varkappa }.  \label{7c.2.12b}
\end{equation}

This means that the energy spectrum of the s.a. Calogero Hamiltonian $\hat{H}%
_{2,\lambda }$ with $\lambda \geq 0$ is the semiaxis $\mathbb{R}_{+}$, $%
\mathrm{spec\,}\hat{H}_{2,\lambda }=[0,\infty ),\,\lambda \geq 0,$
continuous and simple, as well as the spectrum of $\hat{H}_{1}$.

The generalized eigenfunctions are $u_{2,\lambda }(x;E),\,E\geq 0$, given by
(\ref{7c.2.5}) with the substitution $W=E$ and $\beta =\sqrt{E}$,%
\begin{equation}
\,u_{2,\lambda }(x;E)=\left( E/4k_{0}^{2}\right) ^{-\varkappa
/2}x^{1/2}J_{\varkappa }(\sqrt{E}x)+\tilde{\lambda}\left(
E/4k_{0}^{2}\right) ^{\varkappa /2}x^{1/2}J_{-\varkappa }(\sqrt{E}%
x),\,\,E\geq 0.  \label{E}
\end{equation}

The normalized generalized\textrm{\ }eigenfunctions%
\begin{align}
& u_{2,\lambda ,E}\,\left( x\right) =\frac{1}{\sqrt{2}}\frac{%
x^{1/2}J_{\varkappa }(\sqrt{E}x)+\gamma (\lambda ,E)x^{1/2}J_{-\varkappa }(%
\sqrt{E}x)}{\sqrt{1+2\gamma (\lambda ,E)\cos \pi \varkappa +\gamma
^{2}(\lambda ,E)}},\,  \label{7c.2.14} \\
& \gamma (\lambda ,E)=\lambda \frac{\Gamma (1-\varkappa )}{\Gamma
(1+\varkappa )}\left( E/4k_{0}^{2}\right) ^{\varkappa },\ \,\lambda \geq
0,\,\,E\ \geq 0,  \notag
\end{align}%
form a complete orthonormalized system of eigenfunctions for the s.a.
Calogero Hamiltonian $\hat{H}_{2,\lambda }$ with $\lambda \geq 0$.

For $\lambda =0$, the inversion formulas coincide with the formulas for the
Fourier--Bessel transformation.

2. Let $\lambda <0.$ The only difference from the case $\lambda >0$ is that
the function $\omega _{2,\lambda }(E)$, given by the same Eq. (\ref{D}), now
has a unique simple zero $E_{2,\lambda },\,\ \omega _{2,\lambda
}(E_{2,\lambda })=0,$ in the negative energy region, 
\begin{equation}
E_{2,\lambda }=-4k_{0}^{2}\left\vert \lambda \frac{\Gamma (1-\varkappa )}{%
\Gamma (1+\varkappa )}\right\vert ^{-1/\varkappa },  \label{7c.2.16}
\end{equation}%
and\footnote{%
Both $\widetilde{\lambda }$ and $E_{2,\lambda }$ are negative, and therefore 
$d\omega _{2,\lambda }(E_{2,\lambda })/dE$ is also negative.}%
\begin{equation*}
\frac{d\omega _{2,\lambda }(E_{2,\lambda })}{dE}=-\frac{2\varkappa \sin \pi
\varkappa }{\pi }\frac{\widetilde{\lambda }}{E_{2,\lambda }}.
\end{equation*}%
Therefore, (\ref{7c.2.12a}) changes to%
\begin{equation*}
\sigma ^{\prime }(E)=\frac{\pi }{2\varkappa \sin \pi \varkappa }\frac{%
E_{2,\lambda }}{\widetilde{\lambda }}\delta (E-E_{2,\lambda })+\frac{\theta
(E)}{2\zeta _{2,\lambda }(E)}\left( E/4k_{0}^{2}\right) ^{\varkappa },\
\,\lambda <0,
\end{equation*}%
and $\zeta _{2,\lambda }(E)$ is given by the same relation (\ref{7c.2.12b}),
of course, with $\lambda <0$.\ 

This means that the energy spectrum of the s.a. Calogero Hamiltonian $\hat{H}%
_{2,\lambda }$ with $\lambda <0$ is the union of a discrete spectrum, the
negative energy level $E_{2,\lambda }$ (\ref{7c.2.16}) corresponding to a
bound state, and a continuous spectrum, the semiaxis $\mathbb{R}_{+}$, 
\begin{equation*}
\mathrm{spec\,}\hat{H}_{2,\lambda }=\left\{ -4k_{0}^{2}\left\vert \lambda 
\frac{\Gamma (1-\varkappa )}{\Gamma (1+\varkappa )}\right\vert
^{-1/\varkappa }\right\} \cup \lbrack 0,\infty ),\ \lambda <0.
\end{equation*}

For the bound-state eigenfunction, we take\footnote{%
We change the sign for the sake of convenience.} $u_{_{2,\lambda
}}(x;E_{2,\lambda })=v_{_{2}}(x;E_{2,\lambda })$; the last equality follows
from (\ref{7c.2.6}) with $W=E_{2,\lambda }$: the second term in the r.h.s.
of the last equality in (\ref{7c.2.6}) vanishes because it is proportional
to $\omega _{2,\lambda }(E_{2,\lambda })=0$. Using the second relation in (%
\ref{7c.25ab}) and (\ref{7c.2.6}), we find 
\begin{equation*}
u(x,E_{2,\lambda })=-\frac{2\sin \pi \varkappa }{\pi }|\widetilde{\lambda }%
|^{1/2}x^{1/2}K_{\varkappa }(\sqrt{|E_{2,\lambda }|}x).
\end{equation*}%
The generalized eigenfunctions of the continuous spectrum $u_{2,\lambda
}(x;E)$ are given by (\ref{E}) with $\lambda <0$.

The normalized bound-state eigenfunction%
\begin{align}
& u_{E_{2,\lambda }\,}(x)=-\left( \frac{\pi }{2\varkappa \sin \pi \varkappa }%
\frac{E_{2,\lambda }}{\widetilde{\lambda }}\right) ^{1/2}u(x,E_{2,\lambda })
\notag \\
& =\sqrt{\frac{2\sin \pi \varkappa }{\pi \varkappa }}|E_{2,\lambda
}|^{1/2}x^{1/2}K_{\varkappa }(\sqrt{|E_{2,\lambda }|}x),  \label{7c.2.19a} \\
& E_{2,\lambda }=-4k_{0}^{2}\left\vert \lambda \frac{\Gamma (1-\varkappa )}{%
\Gamma (1+\varkappa )}\right\vert ^{-1/\varkappa },\ \lambda <0,  \notag
\end{align}%
and the normalized generalized eigenfunctions of the continuous spectrum%
\begin{align}
& u_{2,\lambda ,E}\,\left( x\right) =\frac{1}{\sqrt{2}}\frac{%
x^{1/2}J_{\varkappa }(\sqrt{E}x)+\gamma (\lambda ,E)x^{1/2}J_{-\varkappa }(%
\sqrt{E}x)}{\sqrt{1+2\gamma (\lambda ,E)\cos \pi \varkappa +\gamma
^{2}(\lambda ,E)}},\,  \label{7c.2.19b} \\
& \gamma (\lambda ,E)=\lambda \frac{\Gamma (1-\varkappa )}{\Gamma
(1+\varkappa )}\left( E/4k_{0}^{2}\right) ^{\varkappa },\,\ \lambda
<0,\,\,E\ \geq 0,  \notag
\end{align}%
form a complete orthonormalized system of eigenfunctions for the s.a.
Calogero Hamiltonian $\hat{H}_{2,\lambda }$ with $\lambda <0$.

Apart from the sign of $\lambda$, the inversion formulas and the Parseval
equality for $\hat{H}_{2,\lambda}$ with $\lambda<0$ differ from those for $%
\hat {H}_{2,\lambda}$ with $\,\lambda\geq0$ by an additional term stemming
from a supplementary bound state.

The inversion formulas with a nonzero $\lambda$ are known and can be found,
for example, in \cite{Titch46}, where they are obtained by a different
method.

We now turn to the remaining case of $\left\vert \lambda \right\vert =\infty 
$, i.e., to the s.a. Calogero Hamiltonian $\hat{H}_{2,\infty }$, specified
by a.b. conditions (\ref{6b.7da}) with $\left\vert \lambda \right\vert
=\infty $. Armed with the experience of the preceding consideration, we
restrict ourselves to presenting the main items with short comments.

The a.b. conditions (\ref{6b.7da}) with $\left\vert \lambda \right\vert
=\infty $, formulas (\ref{7c.2.1}) and (\ref{7c.2.2}), on the one hand, and
formulas (\ref{7c.18b}) and (\ref{7c.20b}), representation (\ref{7c.16}),
and relation (\ref{7c.2.4}), on the other hand, define the following choice
for the functions $u,\widetilde{u}$, and $v$ in this case:%
\begin{align}
& u_{2,\infty }(x;W)=u_{2}(x;W)=\left( \beta /2k_{0}\right) ^{\varkappa
}x^{1/2}J_{-\varkappa }(\beta x),  \notag \\
& \widetilde{u}_{2,\infty }(x;W)=u_{1}(x;W)=\left( \beta /2k_{0}\right)
^{-\varkappa }x^{1/2}J_{\varkappa }(\beta x),  \notag \\
& v_{2,\infty }(x;W)=i\sin \pi \varkappa \left( \beta /2k_{0}\right)
^{\varkappa }x^{1/2}H_{\varkappa }^{(1)}(\beta x)  \notag \\
& =u_{2,\infty }(x;W)-e^{-i\pi \varkappa }\left( \beta /2k_{0}\right)
^{2\varkappa }\widetilde{u}_{2,\infty }(x;W),  \label{7c.2.20b}
\end{align}%
with $\widetilde{\omega }_{2,\infty }=-\mathrm{Wr}(u_{2,\infty },\widetilde{u%
}_{2,\infty })=-2\pi ^{-1}\sin \pi \varkappa $, and 
\begin{equation*}
\omega _{2,\infty }(W)=-\mathrm{Wr}(u_{2,\infty },v_{2,\infty
}\,)=-\,e^{-i\pi \varkappa }\left( \beta /2k_{0}\right) ^{2\varkappa }%
\widetilde{\omega }_{2,\infty }=2\frac{\sin \pi \varkappa }{\pi }e^{-i\pi
\varkappa }\left( \beta /2k_{0}\right) ^{2\varkappa }.
\end{equation*}

The last equality in (\ref{7c.2.20b}) is a copy of the required
representation (\ref{7c.16}) with the evident substitutions $v\rightarrow
v_{2,\infty },\,u\rightarrow u_{2,\infty }$,\ $\,\widetilde{u}\rightarrow 
\widetilde{u}_{2,\infty }$, $\,\omega \rightarrow \,\omega _{2,\infty }$, $\,%
\widetilde{\omega }\rightarrow \widetilde{\omega }_{2}$,\thinspace $\
u\rightarrow u_{2,\infty }$.

The guiding functional is given by%
\begin{equation*}
\Phi (\xi ;W)=\int_{0}^{\infty }u_{2,\infty }(x;W)\,\xi (x)dx,\;\xi \subset
D_{H_{2,\infty }}\cap D_{r}(\mathbb{R}_{+}).
\end{equation*}%
It is simple, which is proved as in the previous subsec. 4.2.

It follows that the spectral density is given by $\sigma ^{\prime }(E)=\pi
^{-1}\lim_{\varepsilon \rightarrow +0}\func{Im}\omega _{2,\lambda
}^{-1}(E+i\varepsilon ).$ The function $\omega _{2,\infty }(E)=\omega
_{2,\infty }(E+i0)$ is given by%
\begin{equation*}
\omega _{2,\infty }(E)=2\frac{\sin \pi \varkappa }{\pi }\left\{ 
\begin{array}{l}
\left( -E/4k_{0}^{2}\right) ^{\varkappa },\,\ E<0, \\ 
e^{-i\pi \varkappa }\left( E/4k_{0}^{2}\right) ^{\varkappa },\,\ E\geq 0.%
\end{array}%
\right.
\end{equation*}%
It is a nonzero real function for $E<0$ and a nonzero complex-valued
function for $E>0$, which allows applying formulas (\ref{7c.17b}) and (\ref%
{7c.17c}) with $\omega (E)=$ $\omega _{2,\infty }(E)$ for evaluating the
spectral density $\sigma ^{\prime }(E)$, and we find that\ $\sigma ^{\prime
}(E)=2^{-1}\theta (E)\left( E/4k_{0}^{2}\right) ^{-\varkappa }.$

This means that the energy spectrum of the s.a. Calogero Hamiltonian $\hat{H}%
_{2,\infty }$ is the semiaxis $\mathbb{R}_{+}$, $\mathrm{spec}\hat{H}%
_{2,\infty }=[0,\infty ),$ continuous and simple.

The generalized eigenfunctions are $u_{2,\infty }(x;E),\ E\geq 0$.

The normalized eigenfunctions 
\begin{equation*}
u_{2,\infty ;E\,}(x)=\frac{1}{\sqrt{2}}x^{1/2}J_{-\varkappa }(\sqrt{E}x),\,\
E\geq 0,
\end{equation*}%
form a complete orthonormalized system of eigenfunctions for the s.a.
Calogero Hamiltonian $\hat{H}_{2,\infty }$.

We note that, with the evident changes $u_{2,\lambda }\rightarrow \lambda
^{-1}u_{2,\lambda }$ and $v_{2}\rightarrow \lambda ^{-1}v_{2}$, all the
results can be obtained from the previous results for $\left\vert \lambda
\right\vert $ $<\infty $ by the formal passage to the limit $\left\vert
\lambda \right\vert \rightarrow \infty $.

The respective inversion formulas coincide with the formulas for the
Fourier--Bessel transformation that is known for the indices of the Bessel
functions larger than $-1$ and do not hold for the indices equal to or less
than $-1$.

Concluding this subsection, we make some remarks for physicists.

It is interesting to note that for $\lambda \geq 0$\ and $\ \left\vert
\lambda \right\vert =\infty $ there is no bound states even if the coupling
constant $\alpha $ is negative, $-1/4<\alpha <0$, so that the Calogero
potential is attractive, while for any finite $\lambda <0$, a single bound
state exists even if $\alpha $ is nonnegative, $0\leq \alpha <3/4$, so that
the Calogero potential is zero or repulsive, and as $\lambda $ changes in
the interval $(-\infty ,0)$, the bound-state energy $E_{2,\lambda \text{ }}$
ranges between $0$ and $-\infty $. If the Calogero Hamiltonian is treated as
the s-wave radial Hamiltonian for the three-dimensional motion, we can
suggest that these phenomena may be interpreted as a manifestation of $%
\delta $-like potentials at the origin. In addition, we emphasize that an
s.a. Hamiltonian with $\alpha =0$, treated as a QM Hamiltonian for a free
motion of a nonrelativistic particle on a semiaxis, is not uniquely defined:
there exists a $U(1)$-family of such Hamiltonians specified by different%
{\Large \ }a.b. conditions at the origin. In particular, a negative energy
level $E_{2,\lambda }$, $\,\lambda <0$, can be treated as a Tamm level: see 
\cite{Tamm33}.

\subsection{Third region: $\protect\alpha =-1/4\,(\varkappa =0)$}

For $\alpha =-1/4$, there exists a one-parameter $U\left( 1\right) $-family
of s.a. Calogero Hamiltonians $\hat{H}_{3,\lambda }$,\ $|\lambda |\leq
\infty ,$ defined on the domains $D_{H_{3,\lambda }}$ given by (\ref{6b.11}).

The spectral analysis for the Hamiltonian $\hat{H}_{3,\lambda }$ is similar
to that for the Hamiltonian $\hat{H}_{2,\lambda }$ based on representations (%
\ref{7c.16}) and (\ref{7c.17}) and presented in detail in the previous
subsection. We will therefore dwell only on specific features of the case.

A specific feature of the case under consideration is that the functions $%
u_{1}(x;W)\,$(\ref{7c.18a}) and $u_{2}(x;W)$ (\ref{7c.2.1}) with $%
\varkappa=0 $ coincide, and therefore, we have to find an alternative to the
function $u_{2}(x;W)$.

We note that the function $x^{1/2}N_{0}(\beta x)$ is a solution of equation (%
\ref{7c.1}) with $\alpha =-1/4$ linearly independent of the solution $%
u_{1}(x;W)=x^{1/2}J_{0}(\beta x)$ and recall that the relation 
\begin{equation}
\frac{\pi }{2}N_{0}(z)=\left( \ln z+\mathbf{C}\right) J_{0}(z)-R_{0}(z)
\label{7c.3.1}
\end{equation}%
holds true, where $\mathbf{C}$ is the Euler constant and%
\begin{equation*}
R_{0}(z)=\sum_{k=1}^{\infty }\frac{(-1)^{k}}{(k!)^{2}}\left( \frac{z}{2}%
\right) ^{2k}\sum_{m=1}^{k}\frac{1}{m}
\end{equation*}%
is a function real-entire in $z^{2}$. It follows that the function 
\begin{align*}
u_{3}(x;W)& =x^{1/2}\left[ \frac{\pi }{2}N_{0}(\beta x)-\left( \ln \beta
/2k_{0}+\mathbf{C}\right) J_{0}(\beta x)\right] \\
& =x^{1/2}\left[ J_{0}(\beta x)\ln (k_{0}x)-R_{0}(\beta x)\right] ,
\end{align*}%
where $\ln (\beta /2k_{0})=\ln (\sqrt{|W|}/2k_{0})+i\varphi /2$, is a
solution of equation (\ref{7c.1}) with $\alpha =-1/4$ linearly independent
of the solution $u_{1}(x;W)$ and is real-entire in $W$, as well as $%
u_{1}(x;W)$. The asymptotic behavior of this function at the origin is given
by%
\begin{equation*}
u_{3}(x;W)=x^{1/2}\ln (k_{0}x)+O(x^{3/2}\ln x),\,\ x\rightarrow 0,
\end{equation*}%
whence it follows in particular, together with (\ref{7c.19a}), $\varkappa =0$%
, that $\mathrm{Wr}(u_{3},u_{1})=-1.$

The a.b. conditions (\ref{6b.10}) send us to distinguish the cases of $%
\left\vert \lambda \right\vert <\infty $ and $\left\vert \lambda \right\vert
=\infty $.

We first consider the case of $\left| \lambda\right| <\infty$.

The a.b. conditions (\ref{6b.10}) with $\left\vert \lambda\right\vert
<\infty $, the known properties of the functions $u_{1}$, $u_{3}$, and $%
v_{1} $ (\ref{7c.18b}) with $\varkappa=0$, representation (\ref{7c.16}) and
the relations $H_{0}^{(1)}(z)=J_{0}(z)+$ $iN_{0}(z)$ and (\ref{7c.3.1})
define the following choice for the functions $u$, $\widetilde{u}$, and $v$
in this case: 
\begin{align}
& u_{3,\lambda }(x;W)=\lambda u_{1}(x,W)+u_{3}(x,W)  \notag \\
& =\lambda x^{1/2}J_{0}(\beta x)+x^{1/2}\left[ \frac{\pi }{2}N_{0}(\beta
x)-\left( \ln \beta /2k_{0}+\mathbf{C}\right) J_{0}(\beta x)\right]  \notag
\\
& =x^{1/2}\left[ \widetilde{\lambda }(W)J_{0}(\beta x)+\frac{\pi }{2}%
N_{0}(\beta x)\right] ,  \label{7c.3.6} \\
& \widetilde{u}_{3}(x;W)=u_{1}(x,W),  \notag \\
& v_{3}\left( x;W\right) =-\frac{i\pi }{2}x^{1/2}H_{0}^{(1)}(\beta x)=x^{1/2}%
\left[ \frac{\pi }{2}N_{0}(\beta x)-\frac{i\pi }{2}J_{0}(\beta x)\right] 
\notag \\
& =u_{3,\lambda }(x,W)-\left( \widetilde{\lambda }(W)+i\frac{\pi }{2}\right) 
\widetilde{u}_{3}(x,W),  \label{7c.3.7}
\end{align}%
where $\widetilde{\lambda }(W)=\lambda -\mathbf{C}-\ln \beta /2k_{0},$ with $%
\widetilde{\omega }_{3}=-\mathrm{Wr}(u_{3,\lambda },\widetilde{u}_{3})=-%
\mathrm{Wr}(u_{3},u_{1})=1$ and 
\begin{align*}
& \omega _{3,\lambda }(W)=-\mathrm{Wr}(u_{3,\lambda },v_{3})=-\left( 
\widetilde{\lambda }(W)+i\frac{\pi }{2}\right) \widetilde{\omega }_{3}= \\
& =-\left( \widetilde{\lambda }(W)+i\frac{\pi }{2}\right) =\ln \beta /2k_{0}+%
\mathbf{C-\lambda }.-i\frac{\pi }{2}.
\end{align*}%
The last equality in (\ref{7c.3.7}) is the required copy of representation (%
\ref{7c.16}) for $v_{3}$.

The guiding functional is given by 
\begin{equation*}
\Phi (\xi ;W)=\int_{0}^{\infty }u_{3,\lambda }(y;W)\xi (y)dy,\ \xi \in
D_{H_{3,\lambda }r}\cap D_{r}(\mathbb{R}_{+}).
\end{equation*}

It is simple, which is proved similarly to subsec. 4.2.

It follows that the spectral density is given by $\sigma ^{\prime }(E)=\pi
^{-1}\lim_{\varepsilon \rightarrow +0}\func{Im}\omega _{3,\lambda
}^{-1}(E+i\varepsilon ).$ The function $\omega _{3,\lambda }(E)=\omega
_{3,\lambda }(E+i0)$ given by 
\begin{equation*}
\omega _{3,\lambda }(E)=\left\{ 
\begin{array}{l}
\frac{1}{2}\ln \left( -E/4k_{0}^{2}\right) +\mathbf{C-\lambda },\ \,E<0, \\ 
\frac{1}{2}\ln \left( E/4k_{0}^{2}\right) +\mathbf{C-\lambda }-i\frac{\pi }{2%
},\ \,E>0,%
\end{array}%
\right.
\end{equation*}%
is real on the negative semiaxis and has a single simple zero $E_{3,\lambda
}=-4k_{0}^{2}\,e^{2(\lambda -\mathbf{C})}$, $\omega _{3,\lambda
}(E_{3,\lambda })=0$, with $\omega _{3,\lambda }^{\prime }(E_{_{3,\lambda
}})=\left( 2E_{_{3,\lambda }}\right) ^{-1},$ while on the semiaxis $\mathbb{R%
}_{+}$, it is a nonzero complex-valued function. This allows using formulas (%
\ref{7c.17b}) and (\ref{7c.17c}) for evaluating the spectral density to yield%
\begin{equation*}
\sigma ^{\prime }(E)=2(-E_{3,\lambda })\delta (E-E_{3,\lambda })+\frac{%
\theta (E)}{2\zeta _{3,\lambda }(E)},
\end{equation*}%
where%
\begin{equation*}
\zeta _{3,\lambda }(E)=\widetilde{\lambda }^{2}(E)+\frac{\pi ^{2}}{4}=\left( 
\frac{1}{2}\ln \left( E/4k_{0}^{2}\right) +\mathbf{C-}\lambda \right) ^{2}+%
\frac{\pi ^{2}}{4}.
\end{equation*}

This means that the simple energy spectrum of the s.a. Calogero Hamiltonian $%
\hat{H}_{3,\lambda }$ is the union of a discrete spectrum, the negative
energy level $E_{3,\lambda }$ corresponding to a bound state, and a
continuous spectrum, the semiaxis $\mathbb{R}_{+}$, $\mathrm{spec\,}\hat{H}%
_{2,\lambda }=\{-4k_{0}^{2}e^{2(\lambda -\mathbf{C})}\}\cup \lbrack 0,\infty
).$

We note that in contrast to the second region of values of $\alpha $, $%
-1/4<\alpha <3/4$, a bound state in the case of $\alpha =-1/4$ exists for
any finite $\lambda $, and, as $\lambda $ changes in the interval $(-\infty
,\,\infty )$, the bound-state energy $E_{3,\lambda }$ ranges between $0$ and 
$-\infty $. In some sense, it is misterious if we treat the Calogero
Hamiltonian with $\alpha =-1/4$ as an s-wave radial Hamiltonian for a free
particle in two dimensions and the extension parameter $\lambda $ as a
manifestation of a $\delta $-like potential that can have any sign.

The normalized bound-state eigenfunction%
\begin{align*}
& u_{E_{3,\lambda }}(x)=-\sqrt{2}|E_{3,\lambda }|^{1/2}\,u_{3,\lambda
}(x;E_{3,\lambda })=-\sqrt{2}|E_{3,\lambda }|^{1/2}v_{3}(x;E_{3,\lambda }) \\
& =\sqrt{2}|E_{3,\lambda }|^{1/2}\,x^{1/2}K_{0}(\sqrt{|E_{3,\lambda }|}%
x),\,\,E_{3,\lambda }=-4k_{0}^{2}\,e^{2(\lambda -\mathbf{C})},
\end{align*}%
and the normalized generalized eigenfunctions of the continuous spectrum 
\begin{align*}
& u_{3,\lambda ,\,E}\,(x)=\frac{1}{\sqrt{2\,\zeta _{3,\lambda }(E)}}x^{1/2}%
\left[ \widetilde{\lambda }(E)J_{0}(\sqrt{E}x)+\frac{\pi }{2}N_{0}(\sqrt{E}x)%
\right] , \\
\,& \zeta _{3,\lambda }(E)=\widetilde{\lambda }^{2}(E)+\frac{\pi ^{2}}{4}%
,\,\,\widetilde{\lambda }(E)=\lambda -\mathbf{C}-\frac{1}{2}\ln \left(
E/4k_{0}^{2}\right) ,\,\,E\geq 0,
\end{align*}%
form a complete orthonormalized system of eigenfunctions for the s.a.
Calogero Hamiltonian $\hat{H}_{2,\infty }$.

As to the remaining case of $\left| \lambda\right| =\infty$, we only outline
the main points.

The functions $u_{3,\lambda }$ (\ref{7c.3.6}) and $v_{3}$ (\ref{7c.3.7}) in
the previous case of $\left\vert \lambda \right\vert <\infty $\thinspace are
evidently replaced by the respective functions%
\begin{equation*}
u_{3,\infty }(x;W)=u_{1}(x,W)=x^{1/2}J_{0}(\beta x)
\end{equation*}%
and%
\begin{equation*}
v_{3,\infty }(x,W)=\frac{x^{1/2}H_{0}^{(1)}(\beta x)}{1+\frac{2i}{\pi }%
\left( \ln \beta /2k_{0}+\mathbf{C}\right) }=u_{1}(x,W)+\frac{u_{3}(x;W)}{%
\left( \ln \beta /2k_{0}+\mathbf{C}\right) -\frac{i\pi }{2}}.
\end{equation*}

The guiding functional given by%
\begin{equation*}
\Phi (\xi ;W)=\int_{0}^{\infty }u_{1}(y;W)\xi (y)dy,\;\xi \subset
D_{H_{3,\infty }}\cap D_{r}(\mathbb{R}_{+})
\end{equation*}%
is simple; the proof is similar to that in subsec. 4.2.

The spectral density is $\sigma ^{\prime }(E)=2^{-1}\theta (E).$ This means
that the spectrum of the s.a. Calogero Hamiltonian $\hat{H}_{3,\infty }$ is
the semiaxis $\mathbb{R}_{+}$, $\mathrm{spec\,}\hat{H}_{3,\infty
}=[0,\infty) $; it is continuous and simple.

The normalized generalized eigenfunctions 
\begin{equation*}
u_{3,\infty ;E}\,(x)=\frac{1}{\sqrt{2}}x^{1/2}J_{0}(\sqrt{E}x)\,,\,E\geq 0,
\end{equation*}%
form a complete orthonormalized system of eigenfunctions for the s.a.
Calogero Hamiltonian $\hat{H}_{3,\infty }$.

With the evident changes $u_{3,\lambda }\rightarrow \lambda
^{-1}u_{3,\lambda }$ and $v_{3}\rightarrow \lambda ^{-1}v_{3}$, all the
results can be obtained from the previous results for $\left\vert \lambda
\right\vert $ $<\infty $ by the formal passage to the limit $\left\vert
\lambda \right\vert \rightarrow \infty $.

The respective inversion formulas coincide with the standard formulas for
the Fourier--Bessel transformation.

\subsection{Fourth region: $\protect\alpha <-1/4\,(\varkappa =i\protect%
\sigma ,\,\protect\sigma >0)$}

For each $\alpha$ in this region, there exists a one-parameter $U\left(
1\right) $-family of s.a. Calogero Hamiltonians $\hat{H}_{4,\theta}$, $%
0\leq\theta\leq\pi$, $\theta=0\sim\theta=\pi$, defined on the domains $%
D_{H_{4,\theta}}$ given by (\ref{6b.15}).

The spectral analysis for the Hamiltonian $\hat{H}_{4,\theta }$ is
completely similar to that for the Hamiltonians $\hat{H}_{2,\lambda }$ and $%
\hat{H}_{3,\lambda }$ in the previous two subsections based on
representations (\ref{7c.16}) and (\ref{7c.17}), and we therefore
concentrate only on distinctive features of the case.

The first distinctive feature is that the two linearly-independent solutions 
$u_{1}$ (\ref{7c.18a}) and $u_{2}$ (\ref{7c.2.1}) of equation (\ref{7c.1})
with $\alpha <-1/4$ $(\varkappa =i\sigma )$ are no longer real-entire: they
are entire in $W$, but complex-conjugate on the real axis, $u_{2}(x;E)=%
\overline{u_{1}\left( x;E\right) }$. Relevant linearly-independent
real-entire solutions are 
\begin{align*}
u_{+,\vartheta }\,(x,W)& =e^{i\vartheta }u_{1}(x,W)+e^{-i\vartheta
}u_{2}(x,W),\, \\
u_{-,\vartheta }(x,W)\,& =i\left( e^{-i\vartheta }u_{2}(x,W)-e^{i\vartheta
}u_{1}(x,W)\right) ,
\end{align*}%
with $\mathrm{Wr}(u_{+,\vartheta },u_{-,\vartheta })=2i\mathrm{Wr}%
(u_{1},u_{2})=4\pi ^{-1}\sinh \pi \varkappa ,$ for any fixed$\,\vartheta =%
\overline{\vartheta }$ considered $\func{mod}2\pi $. Accordingly,%
\begin{align*}
u_{1}(x,W)& =\frac{1}{2}e^{-i\vartheta }\left[ u_{+,\vartheta
}\,(x,W)+i\,u_{-,\vartheta }(x,W)\right] , \\
u_{2}(x,W)& =\frac{1}{2}e^{i\vartheta }\left[ u_{+,\vartheta
}\,(x,W)-i\,u_{-,\vartheta }(x,W)\right] .
\end{align*}

For the functions $u,\widetilde{u}$, and $v$ in this case, we take%
\begin{align}
& u_{4,\theta }(x,W)=u_{+,\tilde{\theta}}(x,W)=  \notag \\
& =e^{i\tilde{\theta}}\left( \beta /2k_{0}\right) ^{-i\sigma
}x^{1/2}J_{i\sigma }(\beta x)+e^{-i\tilde{\theta}}\left( \beta
/2k_{0}\right) ^{i\sigma }x^{1/2}J_{-i\sigma }(\beta x),  \notag \\
& \widetilde{u}_{4,\theta }(x,W)=u_{-,\tilde{\theta}}(x,W)=  \notag \\
& =i\left[ e^{-i\tilde{\theta}}\left( \beta /2k_{0}\right) ^{i\sigma
}x^{1/2}J_{-i\sigma }(\beta x)-e^{i\tilde{\theta}}\left( \beta
/2k_{0}\right) ^{-i\sigma }x^{1/2}J_{i\sigma }(\beta x)\right] ,  \notag \\
& v_{4,\theta }(x,W)=\frac{2\sinh \pi \sigma }{e^{\pi \sigma }e^{-i\tilde{%
\theta}}\left( \beta /2k_{0}\right) ^{i\sigma }-e^{i\tilde{\theta}}\left(
\beta /2k_{0}\right) ^{-i\sigma }}\,x^{1/2}\,H_{i\sigma }(\beta x)  \notag \\
& =u_{\theta }(x,W)+i\frac{e^{\pi \sigma }e^{-i\tilde{\theta}}\left( \beta
/2k_{0}\right) ^{i\sigma }+e^{i\tilde{\theta}}\left( \beta /2k_{0}\right)
^{-i\sigma }}{e^{\pi \sigma }e^{-i\tilde{\theta}}\left( \beta /2k_{0}\right)
^{i\sigma }-e^{i\tilde{\theta}}\left( \beta /2k_{0}\right) ^{-i\sigma }}%
\widetilde{u}_{\theta }(x,W),  \label{7c.4.3c}
\end{align}%
where%
\begin{equation*}
\tilde{\theta}=\theta +\theta _{\sigma },\ \ \theta _{\sigma }=\frac{1}{2i}%
\ln \frac{\Gamma (1+i\sigma )}{\Gamma (1-i\sigma )},
\end{equation*}%
with $\widetilde{\omega }_{4}=-\mathrm{Wr}(u_{4,\theta },\widetilde{u}%
_{4,\theta })=-4\pi ^{-1}\sinh \pi \sigma $ and 
\begin{equation}
\omega _{4,\theta }(W)=-\mathrm{Wr}(u_{4,\theta },v_{4,\theta })=-i4\pi
^{-1}\sinh \pi \sigma \frac{e^{\pi \sigma }e^{-i\tilde{\theta}}\left( \beta
/2k_{0}\right) ^{i\sigma }+e^{i\tilde{\theta}}\left( \beta /2k_{0}\right)
^{-i\sigma }}{e^{\pi \sigma }e^{-i\tilde{\theta}}\left( \beta /2k_{0}\right)
^{i\sigma }-e^{i\tilde{\theta}}\left( \beta /2k_{0}\right) ^{-i\sigma }}. 
\notag
\end{equation}%
It is easy to see or check that these functions possess all the required
properties; in particular, the last equality in (\ref{7c.4.3c}), being a
copy of representation (\ref{7c.16}), follows from relation (\ref{7c.2.4})
with $\varkappa =i\sigma $.

The guiding functional given by%
\begin{equation*}
\Phi (\xi ;W)=\int_{0}^{\infty }u_{4,\theta }(y;W)\xi (y)dy,\;\xi \subset
D_{H_{4,\theta }}\cap D_{r}(\mathbb{R}_{+})
\end{equation*}%
$\ $is simple, which is proved similarly to subsec. 4.2.

It follows that the spectral density is given by $\sigma ^{\prime }(E)=\pi
^{-1}\lim_{\varepsilon \rightarrow +0}\func{Im}\omega _{4,\theta
}^{-1}(E+i\varepsilon ).$ The function $\omega _{4,\theta }(E)=\omega
_{4,\theta }(E+i0)$ is given by%
\begin{equation*}
\omega _{4,\theta }(E)=-\frac{4\sinh \pi \sigma }{\pi }\left\{ 
\begin{array}{l}
\cot \frac{1}{2}\Phi _{\theta }(-E),\ \,E<0, \\ 
i\frac{e^{\pi \sigma }+e^{-i\Phi _{\theta }(E)}}{e^{\pi \sigma }-e^{-i\Phi
_{\theta }(E)}},\ \,E>0,%
\end{array}%
\right.
\end{equation*}%
where$\,\Phi _{\theta }(E)=\sigma \ln (E/4k_{0}^{2})-2\widetilde{\theta }$
is real for $E>0$. On the negative\ semiaxis, this function is real and has
an infinite sequence $\left\{ E_{\theta ,n},\,n\in \mathbb{Z}\right\} $ of
simple zeroes, $\omega _{4,\theta }(E_{\theta ,n})=0$, 
\begin{equation}
E_{\theta ,n}=-4k_{0}^{2}\exp \left( 2\frac{\pi /2+\tilde{\theta}+\pi n}{%
\sigma }\right) ,\;n\in \mathbb{Z}.  \label{7c.4.10}
\end{equation}%
with%
\begin{equation*}
\omega _{4,\theta }^{\prime }(E_{\theta ,n})=-\frac{2\sigma \sinh \pi \sigma 
}{\pi |E_{\theta ,n}|},
\end{equation*}%
while on the positive semiaxis it is nonzero and complex-valued. A simple
calculation by formulas (\ref{7c.17b}) and(\ref{7c.17c}) then yields 
\begin{equation*}
\sigma ^{\prime }(E)=\sum_{n=-\infty }^{\infty }\frac{\pi |E_{\theta ,n}|}{%
2\sigma \sinh (\pi \sigma )}\delta (E-E_{\theta ,n})+\frac{\theta (E)}{4%
\left[ \cosh \pi \sigma +\cos \Phi _{\theta }(E)\right] }.
\end{equation*}

This means that the simple energy spectrum of the s.a. Calogero Hamiltonian $%
\hat{H}_{4,\theta }$ is the union of a discrete spectrum, the infinite
sequence $\left\{ E_{\theta ,n},\,n\in \mathbb{Z}\right\} $ of negative
energy levels $E_{\theta ,n}$ (\ref{7c.4.10}) corresponding to bound states%
\footnote{%
It is this discrete spectrum that was first presented in \cite{Case50}.},
and a continuous spectrum, the semiaxis $\mathbb{R}_{+}$, 
\begin{equation*}
\mathrm{spec\,}\hat{H}_{4,\theta }=\left\{ -4k_{0}^{2}\exp \left( 2\frac{\pi
/2+\tilde{\theta}+\pi n}{\sigma }\right) ,\;n\in \mathbb{Z}\right\} \cup
\lbrack 0,\infty )\ .
\end{equation*}%
The negative energy levels are concentrated exponentially to zero as $%
n\rightarrow -\infty $ and go exponentially to $-\infty $ as $n\rightarrow
\infty $, so that the energy spectrum for all s.a. Calogero Hamiltonians
with $\alpha <-1/4$ is not bounded from below. The radius of the bound
states go to zero as $n\rightarrow \infty $, which manifests the phenomenon
of a \textquotedblleft fall to the center\textquotedblright .

Accordingly, the normalized bound-state eigenfunctions\footnote{%
The sign factors are introduced for the sake of convenience. We also use the
second relation in (\ref{7c.25ab}) with $\varkappa =i\sigma $.
\par
.}%
\begin{align}
& u_{E_{\theta ,n}}(x)=(-1)^{n+1}\left( \frac{\pi |E_{\theta ,n}|}{2\sigma
\sinh \pi \sigma }\right) ^{1/2}u_{4,\theta }(x;E_{\theta ,n})  \notag \\
& =\left( \frac{2\sinh \pi \sigma \,|E_{\theta ,n}|}{\pi \sigma }\right)
^{1/2}x^{1/2}K_{i\sigma }\left( |E_{\theta ,n}|^{1/2}x\right) \,,  \notag \\
& E_{\theta ,n}=-4k_{0}^{2}\exp \frac{2}{\sigma }\left( \pi /2+\tilde{\theta}%
+\pi n\right) ,\;n\in \mathbb{Z},  \label{7c.4.14a}
\end{align}%
and the normalized generalized eigenfunctions of the continuous spectrum 
\begin{align}
& u_{4,\theta ;E\,}(x)=\sqrt{\rho _{4,\theta }(E)}u_{4,\theta \,}(x;E)=\frac{%
1}{2\sqrt{\cosh \pi \sigma +\cos \Phi _{\theta }(E)}}  \notag \\
& \times \left[ e^{i\tilde{\theta}}\left( E/4k_{0}^{2}\right) ^{-i\sigma
/2}x^{1/2}J_{i\sigma }(\sqrt{E}x)+e^{-i\tilde{\theta}}\left(
E/4k_{0}^{2}\right) ^{i\sigma /2}x^{1/2}J_{-i\sigma }(\sqrt{E}x)\right] , 
\notag \\
& \,\Phi _{\theta }(E)=\sigma \ln E/4k_{0}^{2}-2\tilde{\theta},\ \,E\geq 0,
\label{7c.4.14b}
\end{align}%
form a complete orthonormalized system of eigenfunctions for the s.a.
Calogero Hamiltonian $\hat{H}_{4,\theta }$.

We have not succeeded in finding a respective inversion formulas in
mathematical handbooks. These are an extension of the Fourier--Bessel
transformation to imaginary indices of the Bessel functions.

\section{Fate of scale symmetry}

The scale parameter $k_{0}$, introduced for dimensional reasons, appears to
be significant in s.a. extensions for $\alpha<3/4$: its change $%
k_{0}\rightarrow lk_{0}$ generally changes the extension parameter, which
indicates the breaking of scale symmetry.

From the mathematical standpoint, it is convenient to parameterize s.a.
extensions by a dimensionless parameter, $\lambda $ or $\theta $. However,
from the physical standpoint, it seems more appropriate to convert the two
parameters, the fixed dimensional parameter $k_{0}$ of spatial dimension $%
\mathrm{d}_{k_{0}}=-1$ and the varying dimensionless parameters $\lambda $
and $\theta $ of s.a. extensions, to one dimensional parameter $\mu $ of
spatial dimension $\mathrm{d}_{\mu }=-1$ uniquely parameterizing the
extensions, and the parameter $k_{0}$ no longer enters the description. This
makes evident the spontaneous breaking of the scale symmetry.

As is easily seen from (\ref{6b.7da}), in the case of $-1/4<\alpha <3/4$ and
for $\lambda >0$, this parameter is $\mu =k_{0}\lambda ^{-\frac{1}{%
2\varkappa }}\,,\;0<\mu <\infty .\,$The s.a. Calogero Hamiltonian $\hat{H}%
_{2,\lambda }$ with $\lambda >0$ is now naturally labelled by the subscript $%
\mu $ and an extra subscript $+$ indicating the sign of $\lambda $,$\,\hat{H}%
_{2,\mu ,+}=\,\hat{H}_{2,\lambda }$, $\lambda >0$,\thinspace and is
specified by the a.b. conditions 
\begin{align}
\psi _{2,\mu ,+}(x)& =cx^{1/2}\left[ (\mu x)^{\varkappa }+(\mu
x)^{-\varkappa }\right] +O(x^{3/2}),  \notag \\
\psi _{2,\mu ,+}^{\prime }(x)& =cx^{-1/2}\left[ (1/2+\varkappa )(\mu
x)^{\varkappa }+(1/2-\varkappa )(\mu x)^{-\varkappa }\right]
+O(x^{1/2}),\;x\rightarrow 0\,.  \label{7c.5.2}
\end{align}%
The complete orthonormalized system (\ref{7c.2.14}) of eigenfunctions for
the Hamiltonian $\hat{H}_{2,\mu ,+}$ is presented in terms of the scale
parameter $\mu $ as follows: 
\begin{align}
& u_{2,\mu ,+;E}\left( x\right) =\frac{1}{\sqrt{2}}\frac{x^{1/2}J_{\varkappa
}(\sqrt{E}x)+\gamma _{+}(\mu ,E)\,x^{1/2}J_{-\varkappa }(\sqrt{E}x)}{\sqrt{%
1+2\gamma _{+}(\mu ,E)\cos \pi \varkappa +\gamma _{+}^{2}(\mu ,E)}},\, 
\notag \\
& \gamma _{+}(\mu ,E)=\frac{\Gamma (1-\varkappa )}{\Gamma (1+\varkappa )}%
\left( E/4\mu ^{2}\right) ^{\varkappa },\,\,E\ \geq 0;  \label{7c.5.2a}
\end{align}%
the auxiliary scale parameter $k_{0}$ then disappears.

For $\lambda <0$, the dimensional parameter is $\mu =k_{0}|\lambda |^{-\frac{%
1}{2\varkappa }}$,\ $0<\mu <\infty \,.$ The Hamiltonian $\,\hat{H}%
_{2,\lambda }$ with $\lambda <0$ is now denoted by $\hat{H}_{2,\mu ,-}$: $\,%
\hat{H}_{2,\mu ,-}=\,\hat{H}_{2,\lambda }$, $\lambda <0$, and is specified
by the a. b. conditions 
\begin{align}
\psi _{2,\mu ,-}\left( x\right) & =cx^{1/2}\left[ (\mu x)^{\varkappa }-(\mu
x)^{-\varkappa }\right] +O(x^{3/2}),  \notag \\
\psi _{2,\mu ,-}^{\prime }\left( x\right) & =cx^{-1/2}\left[ (1/2+\varkappa
)(\mu x)^{\varkappa }-(1/2-\varkappa )(\mu x)^{-\varkappa }\right]
+O(x^{1/2}),\;x\rightarrow 0\,.  \label{7c.5.4}
\end{align}%
The single negative energy level representing its discrete spectrum is now
given by%
\begin{equation}
E_{2,\mu ,-}=-4\mu ^{2}\left( \frac{\Gamma \left( 1+\varkappa \right) }{%
\Gamma \left( 1-\varkappa \right) }\right) ^{1/\varkappa }.  \label{7c.5.5}
\end{equation}%
The complete orthonormalized system (\ref{7c.2.19a}), (\ref{7c.2.19b}) of
eigenfunctions for the Hamiltonian $\hat{H}_{2,\mu ,-}$ is written in terms
of the scale parameter $\mu $ as%
\begin{align}
& u_{E_{2,\mu ,-}\,}(x)=\sqrt{\frac{2\sin \pi \varkappa }{\pi \varkappa }}%
\left( -E_{2,\mu ,-}\right) ^{1/2}x^{1/2}K_{\varkappa }(\sqrt{-E_{2,\mu ,-}}%
x),  \label{7c.5.5a} \\
& u_{2,\mu ,+;E}\left( x\right) =\frac{1}{\sqrt{2}}\frac{x^{1/2}J_{\varkappa
}(\sqrt{E}x)+\gamma _{-}(\mu ,E)\,x^{1/2}J_{-\varkappa }(\sqrt{E}x)}{\sqrt{%
1+2\gamma _{-}(\mu ,E)\cos \pi \varkappa +\gamma _{-}^{2}(\mu ,E)}},  \notag
\\
& \gamma _{-}(\mu ,E)=-\frac{\Gamma (1-\varkappa )}{\Gamma (1+\varkappa )}%
\left( E/4\mu ^{2}\right) ^{\varkappa },\,\,E\ \geq 0,  \notag
\end{align}%
We note that the s.a. Calogero Hamiltonian $\hat{H}_{2,\mu ,-}$ is uniquely
determined by a position of the negative energy level.

The exceptional values $\lambda =0$ and $\left\vert \lambda \right\vert
=\infty $ of the extension parameter are naturally included in this scheme
as the respective exceptional values $\mu =\infty $ and $\mu =0$ of the
scale parameter, and in terms of $\mu $ the corresponding Hamiltonians are
respectively denoted by $\hat{H}_{2,\infty }$, $\hat{H}_{2,\mu =\infty }=%
\hat{H}_{2,\lambda =0}$, and $\hat{H}_{2,0}$, $\hat{H}_{2,\mu =0}=\hat{H}%
_{2,\lambda =\infty }$.

As is seen from (\ref{6b.10}), in the case of $\alpha =-1/4$ and for $%
\left\vert \lambda \right\vert <\infty $, the dimensional parameter is $\mu
=k_{0}e^{\lambda }\,,\;0<\mu <\infty \,.$ In terms of $\mu $, the respective
s.a. Calogero Hamiltonian $\hat{H}_{3,\mu }$,$\,\hat{H}_{3,\mu }=\hat{H}%
_{\lambda }$, is specified by a.b. conditions, 
\begin{align}
\psi _{3,\mu }(x)& =cx^{1/2}\ln (\mu x)+O(x^{3/2}),  \notag \\
\psi _{3,\mu }^{\prime }(x)& =cx^{-1/2}\left( \frac{1}{2}\ln (\mu
x)+1\right) +O(x^{1/2}),\,x\rightarrow 0\,.  \label{7c.5.7}
\end{align}%
The single negative energy level representing its discrete spectrum is given
by $E_{3,\mu }=-4\mu ^{2}\exp (-2\mathbf{C}),$ where $\mathbf{C}$ is the
Euler constant; a position of this level uniquely determines the Hamiltonian 
$\hat{H}_{3,\mu }$.

The exceptional values $\lambda =-\infty $ and $\lambda =\infty $ of the
extension parameter $\lambda $, which are equivalent, $-\infty \sim \infty $%
, are naturally included as the respective exceptional values $\mu =\infty $
and $\mu =0$ of the scale parameter $\mu $, which are equivalent, $\infty
\sim 0$. In terms of $\mu $, we let $\hat{H}_{3}$ denote the corresponding
Hamiltonian, $\hat{H}_{3}=\hat{H}_{3,|\lambda |=\infty }$.

As is seen from (\ref{6b.14}), in the case of $\alpha <-1/4$, the
dimensional parameter is 
\begin{equation}
\mu =k_{0}e^{\frac{\theta }{\sigma }},\ \mu _{0}\leq \mu \leq \mu
_{0}\,e^{\pi /\sigma },\ \mu _{0}\sim \mu _{0}e^{\pi /\sigma }
\label{7c.5.8}
\end{equation}%
with some fixed $\mu _{0}>0$. In terms of $\mu $, the respective s.a.
Calogero Hamiltonian $\hat{H}_{4,\mu }$, $\hat{H}_{4,\mu }=\hat{H}_{4,\theta
}$, is specified by a.b. conditions, 
\begin{align}
& \psi _{4,\mu }(x)=cx^{1/2}\left[ \left( \mu x\right) ^{i\sigma }+\left(
\mu x\right) ^{-i\sigma }\right] +O(x^{3/2}),  \notag \\
& \psi _{4,\mu }^{\prime }\left( x\right) =cx^{-1/2}\left[ (1/2+i\sigma
)(\mu x)^{i\sigma }-(1/2-i\sigma )(\mu x)^{-i\sigma }\right]
+O(x^{1/2}),\;x\rightarrow 0\,.  \label{7d}
\end{align}%
The infinite sequence of negative energy levels representing its discrete
spectrum is given by%
\begin{equation}
E_{\mu ,n}=-4\mu ^{2}\exp \frac{\pi +2\theta _{\sigma }}{\sigma }\exp \frac{%
2\pi n}{\sigma },\ \,n\in \mathbb{Z};  \label{7c.5.10}
\end{equation}%
a position of one of negative energy levels in any of the intervals%
\begin{equation*}
(-4\mu _{0}^{2}e^{\frac{\theta \left( \sigma \right) +\pi }{\sigma }}\,e^{%
\frac{2\pi m}{\sigma }},\,-4\mu _{0}^{2}e^{\frac{\theta \left( \sigma
\right) -\pi }{\sigma }\,}e^{\frac{2\pi m}{\sigma }}),\,m\in \mathbb{Z},
\end{equation*}%
uniquely determines the Hamiltonian $\hat{H}_{4,\mu }$. The complete
orthonormalized system (\ref{7c.4.14a}), (\ref{7c.4.14b}) of eigenfunctions
for the Hamiltonian $\hat{H}_{4,\mu }$ is written in terms of the scale
parameter $\mu $ as follows:%
\begin{align}
& u_{E_{\mu ,n}}(x)=\left( \frac{2|E_{\mu ,n}|\sinh (\pi \sigma )}{\pi
\sigma }\right) ^{1/2}x^{1/2}K_{i\sigma }\left( |E_{\mu ,n}|^{1/2}x\right)
\,,  \label{7d.a} \\
& u_{4,\mu ;E\,}(x)=\frac{1}{2\sqrt{\cosh \pi \sigma +\cos \Phi _{\mu }(E)}}
\notag \\
& \times \left[ e^{i\theta _{\sigma }}\left( E/4\mu ^{2}\right) ^{-i\sigma
/2}x^{1/2}J_{i\sigma }(\sqrt{E}x)+e^{-i\theta _{\sigma }}\left( E/4\mu
^{2}\right) ^{i\sigma /2}x^{1/2}J_{-i\sigma }(\sqrt{E}x)\right] ,  \notag \\
\,& \Phi _{\mu }(E)=\sigma \ln (E/4\mu ^{2})-2\theta _{\sigma },\,E\geq 0.
\label{7d.b}
\end{align}

The scale parameter $\mu $, as well as $\mu _{0}$, is evidently defined
modulo the factor $\exp \frac{\pi m}{\sigma }$, $\,m\mathbb{=}\in \mathbb{Z}$%
%
%
%
%
%
%
%
; the a.b. conditions (\ref{7d}) are invariant under the change $\mu
\rightarrow e^{\frac{\pi m}{\sigma }}\mu $; accordingly, the discrete
spectrum (\ref{7c.5.10}) is also invariant under this change, and the same
holds for the normalized eigenfunctions (\ref{7d.a}), (\ref{7d.b}) up to the
irrelevant factor $-1$ in front of eigenfunctions of continuous spectrum for
odd $m$.

All s.a. Calogero Hamiltonians that form a $U(1)$-family for each value of
the coupling constant $\alpha $ in all three regions of the values of $%
\alpha <3/4$ are thus parametrized by a scale parameter $\mu $, and in the
region $-1/4<\alpha <3/4$ we must distinguish two different subfamilies by
additional indices $+$ or $-$.

We now turn to the problem of the scale symmetry for s.a. Calogero
Hamiltonians. The scale symmetry is associated with the one-parameter group
of unitary scale transformations $\hat{U}\left( l\right) ,\,l>0$, defined by
(\ref{6a.4}). Under a preliminary \textquotedblleft naive\textquotedblright\
treatment of the Calogero problem, see sec. 2, the \textquotedblleft
naive\textquotedblright\ Hamiltonian\ $\hat{H}$ identified with the initial
differential expression (\ref{6a.2}), which has been considered as an s.a.
operator without any reservations about its domain, formally satisfies the
scale symmetry relation (\ref{6a.5}). It is this relation that is a source
of \textquotedblleft paradoxes\textquotedblright\ concerning the spectrum of
the \textquotedblleft naive\textquotedblright\ $\hat{H}$ . One of our duties
is to resolve these paradoxes.

If we extend relation (\ref{6a.5}) to the s.a. Calogero Hamiltonians $\hat{H}%
_{[i]},[i]=1$; $2,\mu ,+$; $2,\mu ,-$; $3,\mu $; $4,\mu $, we must recognize
that this relation is nontrivial because the operators$\;\hat{H}_{\left[ i%
\right] }$ are unbounded, and, in general, their domains $D_{H_{^{\left[ i%
\right] }}}$ change with changing the scale parameter $\mu $ that naturally
changes under scale transformations. The relation 
\begin{equation}
\hat{U}^{-1}\left( l\right) \hat{H}_{[i]}\hat{U}\left( l\right) =l^{-2}\hat{H%
}_{[i]}\Longleftrightarrow \hat{H}_{[i]}\hat{U}\left( l\right) =l^{-2}\hat{U}%
\left( l\right) \hat{H}_{[i]}\,\,  \label{7c.5.12}
\end{equation}%
for the Hamiltonian $\hat{H}_{[i]}$ with a specific $[i]$, if does hold,
implies that, apart from the fact that \textquotedblleft the rule of action
\textquotedblright\ of the operator $\hat{H}_{^{\left[ i\right] }}$ changes
in accordance with (\ref{7c.5.12}), its domain $D_{H_{^{\left[ i\right] }}}$
is invariant under scale transformations:%
\begin{equation}
\hat{U}(l)D_{H_{^{\left[ i\right] }}}=D_{H_{^{\left[ i\right] }}}\,.
\label{7c.5.13}
\end{equation}%
In such a case, we say that the Hamiltonian $\hat{H}_{[i]}$ is
scale-covariant and is of scale dimension $\mathrm{d}_{H_{[i]}}=-2$; in
short, we speak about the scale symmetry of the Hamiltonian $\hat{H}_{[i]}$.
If relation (\ref{7c.5.13}) does not hold, i.e., if the domain $D_{H_{^{%
\left[ i\right] }}}$ of the Hamiltonian $\hat{H}_{[i]}$ is not
scale-invariant, we are forced to speak about the phenomenon of a
spontaneous breaking of scale symmetry for the Hamiltonian $\hat{H}_{[i]}$.

The initial symmetric operator $\hat{H}$ and its adjoint $\hat{H}^{+}$
associated with the differential expression (\ref{6a.2}) and defined on the
respective domains $D_{H}$ (\ref{6b.1}) and $D_{H^{+}}$ (\ref{6b.1a}) are
scale-covariant because both $D_{H}$\ are $D_{H^{+}}$ are evidently
scale-invariant. The s.a. extensions $\hat{H}_{[i]}$ of the scale covariant $%
\hat{H}$ can lose this property. On the other hand, $\hat{H}_{[i]}$ are s.a.
restrictions of $\hat{H}^{+}$, and their domains $D_{H_{^{\left[ i\right]
}}} $ belong to the scale-invariant domain $D_{H^{+}}$, $D_{H_{^{\left[ i%
\right] }}}\subseteq D_{H^{+}}$ Therefore, the scale symmetry of a specific
Hamiltonian $\hat{H}_{[i]}$ is determined by a behavior of the a.b.
conditions specifying this s.a. operator and thus restricting its domain in
comparison with $D_{H^{+}}$ under scale transformations. This behavior is
different for different $[i]$; namely, it is different for the above four
regions of the values of $\alpha $ (see sec. 3) and strongly depends on the
value of the scale parameter $\mu \,$specifying the s.a. Hamiltonians in
each of the last three regions. We consider these four regions sequentially.

i) First region: $\alpha \geq 3/4.$

For each $\alpha $ in this region, the single s.a. Calogero Hamiltonian $%
\hat{H}_{1}$ coincides with the operator $\hat{H}^{+}$, $\hat{H}_{1}=\hat{H}%
^{+}$, and is therefore scale covariant,%
\begin{equation}
\hat{U}\left( l\right) \hat{H}_{1}\hat{U}^{-1}(l)=l^{-2}\hat{H}_{1.}
\label{7c.5.13a}
\end{equation}%
In other words, the scale symmetry holds for $\alpha \geq 3/4$. The scale
transformation law (\ref{6a.4}) as applied to eigenfunctions (\ref{7c.26b})
yields 
\begin{equation}
u_{1,E}(x)\rightarrow \hat{U}\left( l\right)
u_{1,E}(x)=l^{-1}u_{1,l^{-2}E}(x),  \label{7c.5.14}
\end{equation}%
which we treat, in particular, as the scale transformation law for the
energy spectrum, given by%
\begin{equation}
E\rightarrow l^{-2}E,  \label{7c.5.15}
\end{equation}%
i.e., the spatial dimension of energy $\mathrm{d}_{E}=-2$. The group of
scale transformations acts transitively on the energy spectrum, the semiaxis 
$\mathbb{R}_{+}$, except the point $E=0\ $ that is a stationary point. This
coincides with our preliminary expectations in sec. 2.

ii) Second region: $-1/4<\alpha<3/4.$

The change of a.b. conditions (\ref{7c.5.2}) under scale transformations(\ref%
{6a.4}) is given by the natural scale transformation 
\begin{equation}
\mu \rightarrow l^{-1}\mu ,  \label{7c.5.16a}
\end{equation}%
of the dimensional scale parameter $\mu $ (its spatial dimension being $-1$%
), or, in terms of the dimensionless extension parameter $\lambda $, by 
\begin{equation}
\lambda \rightarrow l^{2\sigma }\lambda ,  \label{7c.5.16b}
\end{equation}%
which implies that under the scale transformations the respective domain $%
D_{H_{2,\mu ,+}}$ of the Hamiltonian $\hat{H}_{2,\mu ,+}$, $0<\mu <\infty $,
transforms to $D_{H_{2,l^{-1}\mu ,+}}$, 
\begin{equation}
D_{H_{2,\mu ,+}}\rightarrow \hat{U}(l)D_{H_{2,\mu ,+}}=D_{H_{2,l^{-1}\mu
,+}}.  \label{7c.5.17}
\end{equation}%
It follows that the scale transformations change the Hamiltonian $\hat{H}%
_{2,\mu ,+}$ to another Hamiltonian $\hat{H}_{2,l^{-1}\mu ,+}$, 
\begin{equation}
\hat{H}_{2,\mu ,+}\rightarrow \hat{U}\left( l\right) \hat{H}_{2,\mu ,+}\hat{U%
}^{-1}(l)=l^{-2}\hat{H}_{2,l^{-1}\mu ,+},  \label{7c.5.18}
\end{equation}%
which means that the scale symmetry is spontaneously broken for the
Hamiltonians $\hat{H}_{2,\mu ,+}$, $0<\mu <\infty $. The scale
transformation law for the eigenfunctions (\ref{7c.5.2a}) is given by%
\begin{equation}
u_{2,\mu ,+;E\,}(x)\rightarrow \hat{U}\left( l\right) u_{2,\mu
,+;E\,}(x)=l^{-1}u_{2,l^{-1}\mu ,+;l^{-2}E}(x),E\geq 0.  \label{7c.5.19}
\end{equation}

The same evidently holds true for the Hamiltonians $\hat{H}_{2,\mu ,-},0<\mu
<\infty $, specified by a.b. conditions (\ref{7c.5.4}): the respective
formulas (\ref{7c.5.16a}) and (\ref{7c.5.16b}) remain unchanged, while in
formulas (\ref{7c.5.17}), (\ref{7c.5.18}), and (\ref{7c.5.19}) the subscript 
$+$ changes to the subscript $-$, and formula (\ref{7c.5.19}) for the
eigenfunctions of the continuous spectrum is supplemented by the formula for
bound-state eigenfunction (\ref{7c.5.5a}), (\ref{7c.5.5}) 
\begin{equation}
u_{\,E_{2,\mu ,-}}\rightarrow \hat{U}\left( l\right) u_{E_{2,\mu
,-}}(x)=u_{E_{2,l^{-1}\mu ,-}}(x),\,E_{2,l^{-1}\mu ,-}=l^{-2}E_{2,\mu ,-}.
\label{7c.5.20}
\end{equation}

The Hamiltonians $\hat{H}_{2,\infty }$ and $\hat{H}_{2,0}$ corresponding to
the respective exceptional values $\mu =\infty $\thinspace $(\lambda =0)$
and $\mu =0\,(\lambda =\infty )$ of the scale parameter $\mu $ and specified
by the respective a.b. conditions 
\begin{equation*}
\psi _{2,\infty }(x)=cx^{1/2+\varkappa }+O(x^{3/2}),\,\ x\rightarrow 0,\ \
\psi _{2,0}(x)=cx^{1/2-\varkappa }+O(x^{3/2}),\,x\rightarrow 0,
\end{equation*}%
are scale-covariant, which means that copies of formulas (\ref{7c.5.13a}), (%
\ref{7c.5.14}), and (\ref{7c.5.15}) with replacing subscript $1$ to the
respective subscripts$\ 2,\infty $ and $2,0$ hold true. If we require scale
symmetry in the Calogero problem, then only the two possibilities, $\hat{H}%
_{2,\infty }$ or $\hat{H}_{2,0}$, remain for the s.a. Calogero Hamiltonian
with $\alpha $, so that $-1/4<\alpha <3/4$. We note that this interval of $%
\alpha $ includes the point $\alpha =0\,(\varkappa =1/2)$ corresponding to a
free motion. Therefore, all the above-said concerning the spontaneous
scale-symmetry breaking relates to the case of a free particle on a semiaxis.

iii) Third region: $\alpha=-1/4.$

The change of the a.b. conditions (\ref{7c.5.7}) under the scale
transformations (\ref{6a.4}) is equivalent to rescaling (\ref{7c.5.16a}) of
the dimensional parameter $\mu $, or to the change $\lambda \rightarrow
\lambda -\ln l$ of the dimensionless extension parameter $\lambda $. A
further consideration is completely similar to the preceding one, to yield
that copies of relations (\ref{7c.5.17}), (\ref{7c.5.18}), (\ref{7c.5.19}),
and (\ref{7c.5.20}), with the subscript $2$ replaced by the subscript $3$,
and with the subscripts $+$ and $-$ eliminated, hold true for the
Hamiltonians $\hat{H}_{3,\mu },0<\mu <\infty $, which implies scale-symmetry
breaking for these Hamiltonians.

As to the Hamiltonian $\hat{H}_{3}$ corresponding to the exceptional values $%
\mu =0$ and $\mu =\infty $ of the scale parameter $\mu $, which are
equivalent, $0\sim \infty $, and specified by the a.b. conditions $\psi
_{3}(x)=cx^{1/2}+O(x^{3/2})$, this Hamiltonian is scale-covariant, and
copies of relations (\ref{7c.5.13a}), (\ref{7c.5.14}), and (\ref{7c.5.15})
with the substitution $1\rightarrow 3$ hold true. If we require scale
symmetry for the s.a. Calogero Hamiltonian with $\alpha $ $=$ $-1/4$, then
it is only the Hamiltonian $\hat{H}_{3}$ that survives.

iv) Fourth region: $\alpha<-1/4.$

The change of the a.b. conditions (\ref{7d}) under the scale transformations
(\ref{6a.4}) is equivalent to a modified rescaling $\mu \rightarrow
l^{-1}\mu \exp \pi m/\sigma ,$ of the dimensional\ extension parameter $\mu $%
, where an integer $m$ is defined by the condition $\mu _{0}\leq l^{-1}\mu
\exp \pi m/\sigma <\mu _{0}\exp \pi /\sigma :$ the changed $\mu $ must
remain within the interval $[\mu _{0},\mu _{0}\exp \pi /\sigma )$, see (\ref%
{7c.5.8}); this is equivalent to the change $\theta \rightarrow \left.
(\theta +\sigma \ln l)\right\vert _{\func{mod}\pi }$ of the dimensionless
extension parameter $\theta $. It follows that for the Hamiltonians $\hat{H}%
_{4,\mu }$, $\mu _{0}\leq \mu \leq \mu _{0}e^{\pi /\sigma }$, $\,\mu
_{0}\sim \mu _{0}\,e^{\pi /\sigma }$, the relations 
\begin{align*}
& D_{H_{4,\mu }}\rightarrow \hat{U}(l)D_{H_{4,\mu }}=D_{H_{4,\mu l^{-1}\exp
\pi m/\sigma }}, \\
& \hat{H}_{4,\mu }\rightarrow \hat{U}\left( l\right) \hat{H}_{4,\mu }\hat{U}%
^{-1}(l)=l^{-2}\hat{H}_{_{4,}\mu l^{-1}\exp \pi m/\sigma }\ , \\
& u_{E_{\mu ,n}}(x)\rightarrow \hat{U}\left( l\right) u_{E_{\mu
,n}}(x)=u_{E_{\mu l^{-1}\exp \pi m/\sigma ,\,n-m}}(x),\,E_{\mu l^{-1}\exp
\pi m/\sigma ,\,n-m}=l^{-2}E_{\mu ,n}, \\
& u_{4,\mu ;E\,}(x)\rightarrow \hat{U}\left( l\right) u_{4,\mu
,;E\,}(x)=(-1)^{m}\,l^{-1}u_{4,\mu l^{-1}\exp \pi m/\sigma ;l^{-2}E}(x)
\end{align*}%
hold true.

This means that the scale symmetry is spontaneously broken for $\hat{H}%
_{4,\mu }$. The peculiar feature of the fourth region is that for $l=\exp
\pi n/\sigma \,,\;n\in \mathbb{Z}$, the scale symmetry holds true. In other
words, the scale symmetry is not broken completely, but to up an infinite
cyclic subgroup. In particular, this subgroup acts transitively on the
discrete energy spectrum.

This is the fate of the scale symmetry in the QM Calogero problem.

The paradoxes concerning the scale symmetry in the Calogero problem and
considered in sec. 2 are thus resolved. Namely, in general, there is no
scale symmetry in the problem for $\alpha <3/4$. In the latter case, the
\textquotedblleft naive\textquotedblright\ Calogero Hamiltonian $\hat{H}$ of
sec. 2 is actually the operator $\hat{H}^{+}$ that is scale-covariant but
not s.a.. As to s.a. Calogero Hamiltonians, all possibilities for a negative
part of the energy spectrum considered in sec. 2 are generally realized by
different Hamiltonians specified by different a.b. conditions. In general,
the scale symmetry shifts energy levels together with Hamiltonians.

We conclude the above consideration with the following remarks for
physicists.

We have a unique QM description of a nonrelativistic particle moving on a
semiaxis in the Calogero potential (\ref{6.0}) with the coupling constant $%
\alpha \geq 3/4$. In the case of $\alpha <3/4$, mathematics presents
different possibilities related to different admissible s.a. asymptotic
boundary conditions at the origin that are specified in terms of the scale
parameter $\mu $. But a final choice, which is reduced to a specific choice
of the scale parameter $\mu $, belongs to the physicist.

The origin of this parameter presents a physical problem, as well as the
physical interpretation of the chosen s.a. Hamiltonian, as a whole. We can
only note that the usual regularization (\ref{6a.3}) of the Calogero
potential by a cut-off at a finite radius and the consequent passage to the
limit of zero radius yields $\mu =\infty $ in the case of $-1/4\leq \alpha
<3/4$; a peculiar feature of the case of $\alpha =-1/4\ $is that $\mu
=\infty $ is equivalent to $\mu =0$. Such a choice of the scale parameter
corresponds to the minimum possible singularity of wave functions, including
eigenfunctions, at the origin. In the case of $\alpha <-1/4$, the
regularization procedure does not provide any answer: the zero-radius limit
does not exist. A suggestion on the nature of the scale parameter $\mu $, $%
0\leq \mu <\infty $, in the case of $-1/4<\alpha <3/4$, $0<\mu <\infty $ in
the case of $\alpha =$ $-1/4$, and $\mu _{0}\leq \mu \leq \mu _{0}\exp \pi
/\sigma $ in the case of $\alpha <-1/4$, has been presented above in sec. 3:
it is conceivable that this parameter is a manifestation of an additional $%
\delta $-like term in the potential.

When deciding on a specific value of the scale parameter $\mu $, one of the
additional arguments can be related to scale symmetry. In the case of $%
\alpha \geq 3/4$, scale symmetry holds true. In the case of $-1/4\leq \alpha
<3/4$, scale symmetry is spontaneously broken for a generic $\mu $. As for
any spontaneously broken symmetry, scale symmetry does not disappear but
transforms one physical system to another nonequivalent physical system. But
if we require scale symmetry, as we do in similar situations with rotation
symmetry or reflection symmetry, then a possible choice strongly narrows to $%
\mu =\infty $ (the minimum possible singularity of wave functions at the
origin) or $\mu =0$ (the maximum possible singularity) in the case of $%
-1/4<\alpha <3/4$ and to $\mu =\infty \sim $ $\mu =0$ (the minimum possible
singularity) in the case of $\alpha =-1/4$. For strongly attractive Calogero
potentials with $\alpha <-1/4$, the requirement of scale symmetry cannot be
fulfilled: scale symmetry is spontaneously broken for any $\mu $.

\subparagraph{Acknowledgement}

Gitman is grateful to the Brazilian foundations FAPESP and CNPq for
permanent support; Tyutin thanks FAPESP and RFBR, grant 08-02-01118; Tyutin
and Voronov thank LSS-1615.2008.2 for partial support.

\end{document}